\documentclass[aps,prb,twocolumn,superscriptaddress]{revtex4-1}

\usepackage{amsmath}
\usepackage{amsfonts}
\usepackage{amssymb}
\usepackage{graphicx}
\usepackage[utf8]{inputenc}
\usepackage[unicode, hidelinks=true]{hyperref}
\usepackage{color}
\usepackage{braket}
\usepackage{dsfont}
\usepackage{MnSymbol}
\usepackage{xspace}
\usepackage{bbold}
\usepackage{diagbox}
\usepackage{setspace}

\setcitestyle{square,numbers}

\newcommand{\CV}[0]{C_\text{V}}

\newcommand{\ie}[0]{i.e.\@\xspace}
\newcommand{\eg}[0]{e.g.\@\xspace}

\newcommand{\cf}[0]{cf.\@\xspace}

\newfont{\tensy}{cmsy10}
\newcommand{\chem}[1]{{$\fontdimen16\tensy=3.0pt
    \fontdimen17\tensy=3.0pt \mathrm{#1}$}}

\newcommand{\ham}[1]{\hat{H}_{#1}}

\renewcommand{\S}[0]{\mathcal{S}}

\newcommand{\fan}[1]{\hat{c}^{\vphantom\dagger}_{#1}}
\newcommand{\fcr}[1]{\hat{c}^{\dagger}_{#1}}

\newcommand{\ban}[1]{\hat{b}^{\vphantom\dagger}_{#1}}
\newcommand{\bcr}[1]{\hat{b}^{\dagger}_{#1}}

\newcommand{\rhoan}[1]{\hat{\rho}^{\vphantom\dagger}_{#1}}

\newcommand{\Q}[1]{\hat{Q}_{#1}}

\newcommand{\fcohan}[1]{c_{#1}}
\newcommand{\fcohcr}[1]{\bar{c}_{#1}}

\newcommand{\rhocohan}[1]{\rho_{#1}}

\newcommand{\Pp}{P_{\! +}}
\newcommand{\Pm}{P_{\! -}}
\newcommand{\Ppm}{P_{\! \pm}}

\newcommand{\Pnoq}{P}
\newcommand{\Ppmbar}{\bar{P}_{\!\pm}}
\newcommand{\Pmbar}{\bar{P}_{\! -}}
\newcommand{\Ppbar}{\bar{P}_{\! +}}

\newcommand{\omz}{\omega_0}

\newcommand{\kF}{k_{\text{F}}}

\newcommand{\kB}{k_{\text{B}}}

\newcommand{\vF}{v_{\text{F}}}

\newcommand{\im}{\mathrm{i}}

\newcommand{\Eeph}{E_{\mathrm{ep}}}

\newcommand{\Eel}{E_{\mathrm{el}}}
\newcommand{\Eph}{E_{\mathrm{ph}}}

\newcommand{\Hc}{\mathrm{H.c.}}
\newcommand{\absolute}[1]{\left| #1 \right|}
\newcommand{\expv}[1]{\left\langle #1 \right\rangle}
\newcommand{\expvtext}[1]{\langle #1 \rangle}

\newcommand{\expvcn}[1]{\left\llangle #1 \right\rrangle_{C_n}}

\begin{document}

\title{Thermal and quantum lattice fluctuations in Peierls chains}

\author{Manuel Weber}
\affiliation{\mbox{Institut f\"ur Theoretische Physik und Astrophysik,
Universit\"at W\"urzburg, 97074 W\"urzburg, Germany}}
\affiliation{\mbox{Department of Physics, Georgetown University, Washington, DC 20057, USA}}
\author{Fakher F. Assaad}
\affiliation{\mbox{Institut f\"ur Theoretische Physik und Astrophysik,
Universit\"at W\"urzburg, 97074 W\"urzburg, Germany}}
\author{Martin Hohenadler}
\affiliation{\mbox{Institut f\"ur Theoretische Physik und Astrophysik,
Universit\"at W\"urzburg, 97074 W\"urzburg, Germany}}

\date{\today}

\begin{abstract}
The thermodynamic and spectral properties of electrons coupled to quantum phonons are studied
within the spinless Holstein model. Using quantum Monte Carlo simulations, we obtain
accurate results for the specific heat and the compressibility, covering the
entire range of electron-phonon couplings and phonon frequencies.
 To this end, we derive an efficient estimator for the
specific heat using the properties of the perturbation expansion.
This allows us to quantitatively test the predictions of Tomonaga-Luttinger liquid theory
 as well as the widely used adiabatic approximation for low phonon frequencies.
A comparison with the spectral functions of electrons and phonons reveals that
the formation of polaron excitations as well as the renormalization of the phonon
mode across the Peierls transition have a pronounced effect on the specific heat in the
adiabatic regime. 
\end{abstract}


\maketitle

\section{Introduction}

In one-dimensional (1D) systems, the Peierls
instability \cite{Frohlich54,Peierls55} can drive a metal-insulator 
transition to a state with long-range charge-density-wave (CDW) order
accompanied by a periodic lattice modulation.  Experimental realizations of
this phenomenon include quasi-1D materials such as
TTF-TCNQ \cite{PhysRevLett.88.096402} or \chem{K_{0.3}MoO_3} \cite{TRAVAGLINI1983289}. A closely related problem is the
spin-Peierls transition \cite{PhysRevB.10.4637} in, \eg, \chem{CuGeO_3} \cite{PhysRevLett.70.3651}.
Even if the competing electron-electron interaction is neglected, a reliable
quantum-phonon description of such transitions remains challenging.

The specific heat $\CV$ is a key experimental probe for the Peierls
transition. While containing less information than, \eg, excitation
spectra, it is easy to measure and exhibits a peak or an anomaly at $T_c$
\cite{PhysRevLett.32.769,WEI1977595,PhysRevLett.57.1907,PhysRevLett.75.771,PhysRevB.58.R2937, PhysRevLett.65.365}.
Some compounds display a mean-field like step in $\CV$, whereas in other
compounds the anomaly is strongly smeared out by fluctuations \cite{Pouget2016332}. The specific heat is
often dominated by phonon contributions and played a key
role for the development of a quantum theory of lattice fluctuations
\cite{AshcroftMermin}. Despite its experimental importance, accurate results for
microscopic models with quantum phonons are rare.

Investigations of 1D models are motivated by the typically much larger intrachain
interactions compared to interchain interactions, as evinced by the fact that
$T^{}_c$ is typically much smaller than the predicted mean-field value
\cite{Pouget2016332}. While interchain couplings are necessary for the
observed finite-temperature phase transition \cite{Pouget2016332}, the physics is determined by the 1D
chains above a crossover temperature. In 1D, the second-order phase
transition at $T_c>0$ is replaced by a crossover, with long-range order only
at $T=0$. The widely used mean-field approaches \cite{Frohlich54,Kuiper55} are exact for classical
phonons ({\it{adiabatic limit}}) at $T=0$. In the latter, thermal fluctuation
effects (including solitons \cite{RevModPhys.60.781}) can be captured qualitatively with fluctuating gap models
\cite{PhysRevB.6.3409,PhysRevLett.31.462} and quantitatively with Monte Carlo
simulations \cite{PhysRevB.94.155150}. The specific heat has a
characteristic peak at a temperature where coherent
CDW correlations and a well-defined Peierls gap emerge \cite{PhysRevB.94.155150}.

Important insights into the effects of quantum lattice fluctuations on Peierls
insulators have come from numerical methods such as exact diagonalization
\cite{PhysRevB.58.13526,PhysRevLett.81.3956,PhysRevB.73.245120}, quantum
Monte Carlo (QMC) \cite{PhysRevB.27.1680,PhysRevB.27.4302,PhysRevB.53.9676,PhysRevB.60.12125,PhysRevLett.83.195,PhysRevB.83.115105}, and the
density-matrix renormalization group (DMRG) \cite{PhysRevB.60.7950,PhysRevLett.95.137207,Hager20071380,0295-5075-87-2-27001,PhysRevLett.80.5607}, as
well as analytical and semi-analytical methods
\cite{PhysRevB.29.4230,Trebst2001,PhysRevB.71.045112,PhysRevB.70.214429,Barkim2015}.  
Importantly, quantum fluctuations can significantly renormalize or even
destroy the Peierls state, allowing for a gapless Luttinger liquid phase and
a Peierls quantum phase transition. The phase diagram of the spinless
Holstein model considered here (Fig.~\ref{Fig:PD}) has been determined quite
precisely from DMRG calculations \cite{PhysRevLett.80.5607,0295-5075-87-2-27001}.

Combined descriptions of both quantum and thermal effects are particularly challenging
but well motivated: for example in \chem{CuGeO_3}, the relevant phonon frequencies are
comparable to the spin exchange constant \cite{PhysRevB.54.1105}. Other
materials (\eg, polyacetylene \cite{RevModPhys.60.781}) fall into the adiabatic regime of small but finite phonon
frequency. QMC results of limited quality are available for the spin-Peierls case
for selected parameters \cite{PhysRevB.60.12125,PhysRevB.70.214429}. Finite-temperature DMRG
calculations---successfully carried out for fermionic systems
\cite{PhysRevB.81.075108, PhysRevB.86.155156}---are so far inhibited by the large phonon Hilbert
space. The determination of $\CV$ from QMC simulations is limited by long
autocorrelations \cite{Hohenadler2008}, large fluctuations, and Trotter
discretization errors \cite{PhysRevB.36.3833}. Whereas the thermodynamic Bethe
ansatz is not applicable beyond the classical-phonon limit \cite{PhysRevB.70.214436}, 
the bosonization has been applied to study the effect of the coupling to quantum phonons 
in the Luttinger liquid phase \cite{PhysRevB.36.968}.

Here, we present accurate numerical results for the thermodynamic
properties of a 1D Holstein model across all different parameter regimes:
adiabatic and nonadiabatic, as well as metal and Peierls insulator.
Previous limitations are overcome by combining a recently developed directed-loop
algorithm for retarded interactions \cite{PhysRevLett.119.097401} with
the calculation of bosonic observables from the perturbation expansion
\cite{PhysRevB.94.245138}. In particular, the specific heat is calculated
directly on lattices of $L=162$ sites. Moreover, we obtain results for 
the single-particle spectral functions of electrons and phonons
for previously inaccessible system sizes and temperatures to explain
the observed low-temperature features of $C_V$.

The article is organized as follows: We introduce the Holstein model
in Sec.~\ref{Sec:Model} and outline the QMC method
used in Sec.~\ref{Sec:Method}. Results are presented in
Sec.~\ref{Sec:Results} and we conclude in Sec.~\ref{Sec:Conclusion}.
In the Appendix, we derive a QMC estimator
for the specific heat of the Holstein model.

\section{Model}
\label{Sec:Model}

To isolate the effect of quantum lattice fluctuations on the thermodynamics
of 1D chains,
we consider a minimal theoretical model.
The Hamiltonian of the spinless Holstein model \cite{Ho59a} is given by
\begin{align}\label{eq:holsteinmodel}
\hat{H}
	=
	-t \sum_i \big( \fcr{i} \fan{i+1} + \Hc \big)
	+ \omz \sum_i  \bcr{i} \ban{i}
	+ g \sum_i \Q{i} \hat{\rho}_i \, .
\end{align}
Here, $\fcr{i}, \fan{i}$ ($\bcr{i}, \ban{i}$) create/annihilate an electron (phonon) at lattice site $i$. The Holstein model consists of an electronic hopping
term with amplitude $t$, Einstein phonons with frequency $\omz$, and a local
coupling between the lattice displacement $\Q{i} = (\bcr{i} + \ban{i}) / \sqrt{2M\omz}$ and the
fermion density $\hat{\rho}_{i} = ( \hat{n}_i - 1/2 )$.
In the following, we only consider
the half-filled case with $\langle\hat{\rho}_i\rangle=0$ and define the
dimensionless coupling constant $\lambda = g^2 / (4M\omz^2 t)$.
Here, $M$ is the mass of the harmonic oscillators. We use $t$ as the unit of
energy and set $\hbar=1$.

\begin{figure}[t]
\centering
\includegraphics[width=\linewidth]{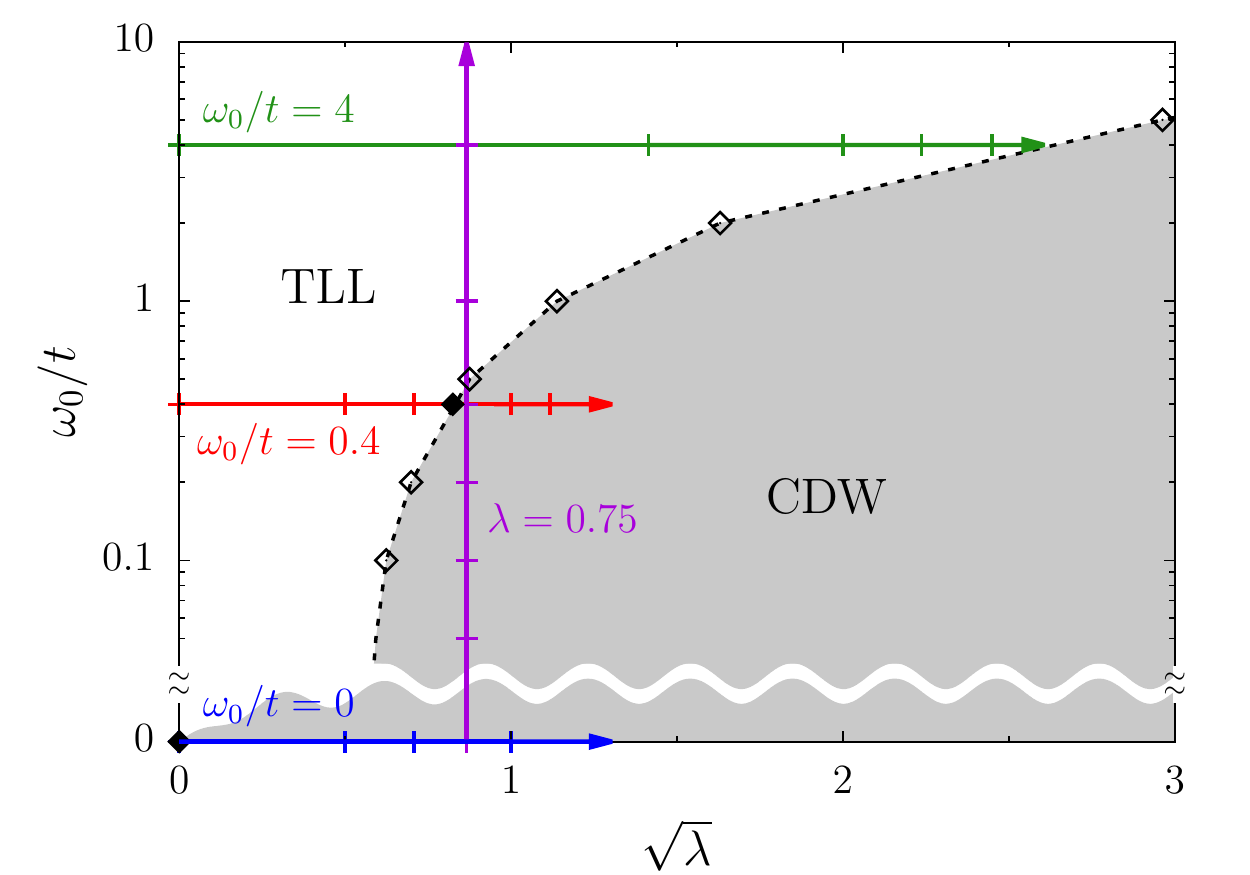}%
\caption{\label{Fig:PD}
Ground-state phase diagram of the half-filled spinless Holstein model
as a function of the electron-phonon coupling $\lambda$ and the phonon
frequency $\omz$. Critical values are
 from DMRG calculations (open symbols) \cite{PhysRevLett.80.5607, 0295-5075-87-2-27001}.
Additionally, we show the QMC estimate $\lambda_c = 0.68(1)$ at $\omz/t=0.4$ \cite{PhysRevLett.119.097401}
and the exact value $\lambda_c = 0$ at $\omz = 0$ (filled symbols).
Arrows indicate the paths in parameter space along which we discuss the
thermodynamic properties of the Holstein model in this article.
The simulation parameters are indicated by additional markers attached to the arrows.
For explanations on the phase diagram see the main text.
}
\end{figure}

Figure \ref{Fig:PD} shows the ground-state phase
diagram of the spinless Holstein model as determined from DMRG simulations
\cite{PhysRevLett.80.5607, 0295-5075-87-2-27001}.
The Holstein model describes the quantum phase transition from
a Tomonaga-Luttinger liquid (TLL) at $\lambda<\lambda_c(\omega_0)$
to a Peierls CDW insulator with ordering
wavevector $q=2\kF=\pi$ at $\lambda>\lambda_c(\omega_0)$ \cite{PhysRevB.27.4302,PhysRevLett.80.5607,PhysRevB.58.13526,PhysRevB.73.245120,0295-5075-87-2-27001}.
At $\omz=0$, the ground state shows CDW order for any
$\lambda >0$ and is exactly described by mean-field theory,
whereas for $\omz > 0$ and small $\lambda$, quantum lattice fluctuations
destroy the ordered state and lead to a TLL phase.
For $\omz \to \infty$, the spinless Holstein model maps to
free fermions and is hence always metallic. The nonuniversal Luttinger
parameters $K$ and $u$ have been determined by DMRG calculations;
for any $\omz$, the electron-phonon interaction leads to a repulsive TLL
with $K<1$ and a reduction of the charge velocity $u$, see
Ref.~\cite{0295-5075-87-2-27001} and references therein.
For further details on the ground-state properties we refer to the review \cite{MHHF2017}.

While the DMRG yields critical values and Luttinger parameters, spectral or
thermodynamic properties appear to be out of reach due to the large bosonic
Hilbert space. Instead, spectral functions have been obtained from
exact diagonalization \cite{PhysRevB.73.245120} or
QMC simulations \cite{PhysRevB.83.115105, PhysRevB.94.245138}
on small system sizes as well as from analytic approaches
\cite{PhysRevB.50.11179, PhysRevB.71.045112, 0295-5075-76-4-644}.
Results for thermodynamic properties are only
available in the adiabatic limit $\omz=0$ \cite{PhysRevB.94.155150}.
The effects of quantum lattice fluctuations have been studied
for a spin-phonon model \cite{PhysRevB.60.12125} but
results are limited by the accessible temperature range and system sizes
because only local QMC updates were available.

The spinless Holstein model captures the essential aspects of quantum Peierls
chains while avoiding complications due to spin gap formation in the metallic
phase that appear in the spinful case \cite{PhysRevB.92.245132}. The phase
transition from TLL to Peierls insulator is also very similar to the
corresponding transition in the spinless Su-Schrieffer-Heeger model \cite{PhysRevB.91.245147}.
Moreover, for classical phonons, the two models exhibit a qualitatively very
similar temperature dependence of the specific heat \cite{PhysRevB.94.155150}. The Jordan-Wigner transformation 
provides a link between the spinless fermion model and spin-phonon models
 \cite{PhysRevB.19.402}. The choice of Einstein
phonons (relevant for, \eg, \chem{CuGeO_3} \cite{PhysRevB.59.14356}) gives
exponential rather than linear (for 1D acoustic phonons) behavior of $\CV$ at low
temperatures. However, a low-energy theory reveals that only the $2\kF$
(zone-boundary) part of the phonon spectrum couples to the electrons
\cite{PhysRevB.27.1680,PhysRevLett.60.2089}, and identical results have been
reported for Su-Schrieffer-Heeger models with optical and acoustic phonons \cite{Barkim2015,PhysRevB.91.245147}. 
Therefore, the decoupled part of the phonon spectrum merely contributes a trivial
background to $\CV$ that is routinely subtracted from experimental data to reveal
the interesting electron-phonon correlation effects.

\section{Method}
\label{Sec:Method}

To simulate the Holstein model, we used the directed-loop QMC method
for retarded interactions in the stochastic series expansion (SSE) representation \cite{PhysRevLett.119.097401}.
Starting from the coherent-state path integral, the phonons are integrated
out analytically \cite{PhysRev.97.660} to obtain the purely fermionic action
\begin{align}
\nonumber
\S
	=
	\S_0
	&- t \int_0^\beta d\tau \sum_{i} \left[ \fcohcr{i}(\tau) \, \fcohan{i+1}(\tau) + \fcohcr{i+1}(\tau) \, \fcohan{i}(\tau) \right] \\
	&- 2 \lambda t \iint_0^\beta d\tau d\tau' \sum_i \rhocohan{i}(\tau) \, P(\tau-\tau') \, \rhocohan{i}(\tau') \, .
\end{align}
The coupling between electrons and phonons leads to a density-density-type
interaction nonlocal in imaginary time and mediated by the free-phonon propagator $P(\tau)$.
The SSE representation \cite{PhysRevB.43.5950}
corresponds to an expansion of the partition function around
$\S_0 = \int d\tau \sum_i \fcohcr{i}(\tau) \, \partial_\tau \, \fcohan{i}(\tau)$.
The resulting trace over Grassmann fields is then mapped to an expectation value of
an operator sequence.
By formally promoting the hopping terms to retarded interactions, we can 
formulate efficient global directed-loop updates from local update rules \cite{Sandvik02}
in which the time dependence of $P(\tau)$ only enters the diagonal updates.
For details on the Monte Carlo updates see Ref.~\cite{PhysRevLett.119.097401}
and its Supplemental Material. Electronic observables are calculated directly
from the Monte Carlo configurations \cite{Sandvik91,PhysRevB.56.14510}.
Bosonic observables can be recovered from electronic correlation functions using sum rules
derived with the help of generating functionals \cite{PhysRevB.94.245138}.

We study the thermodynamics of the Holstein model in terms of the specific heat
\begin{align}
\label{Eq:CV}
C_V
	=
	\kB \beta^2 \left[  \expv{\hat{H}^2} - \expv{\hat{H}}^2 \right]
\end{align}
and the compressibility (we define $\hat{N} = \sum_i \hat{n}_i$)
\begin{align}
\kappa
	=
	\frac{\beta}{L} \left[ \expv{\hat{N}^2} - \expv{\hat{N}}^2 \right] \, .
\end{align}
Here $\beta=1/(\kB T)$ is the inverse temperature.
While $\kappa$ is obtained directly from the world-line configurations,
the calculation of $C_V$ via Eq.~(\ref{Eq:CV}) is complicated
by the fact that the phonon fields have to be extracted from fermionic
correlation functions. In the Appendix, we derive an efficient estimator to
measure $C_V$ in $\mathcal{O}(n)$ operations by exploiting properties of the
interaction expansion ($n$ denotes the expansion order) \cite{PhysRevB.94.245138}.

As a second approach, we also calculated $C_V$ from the
total energy via the relation $C_V = \partial E(T) / \partial T$.
Following Ref.~\cite{PhysRevB.61.9300},
we fitted $E(T)$ to the functional form
\begin{align}
\label{Eq:Ener_MaxEnt}
E(T)
	=
	\int d\omega \, \omega \,
	\big[
		n_\text{F}(\omega,T) \, \rho_{\text{F}}(\omega)
		+ n_\text{B}(\omega,T) \, \rho_{\text{B}}(\omega)
	\big]
\end{align}
which corresponds to a spectral decomposition into
noninteracting fermionic and bosonic contributions $\rho_\text{F}$ and
$\rho_\text{B}$, respectively. This ansatz is well-motivated for the electron-phonon
model at hand---compare Eq.~(\ref{Eq:sum_rule}) and the discussion
of spectral functions below---but
the temperature dependence is considered to
originate only from the Fermi and Bose functions $n_\text{F}$ and $n_\text{B}$.
Given Monte Carlo data for $E(T)$, Eq.~(\ref{Eq:Ener_MaxEnt})
represents an inverse problem that can be solved using the maximum
entropy approach \cite{PhysRevB.61.9300, Jarrell1996133}.
The spectra obtained in this way do not have a physical meaning and
only serve to fit $E(T)$ with a reasonable $\chi^2$.
Then, $C_V(T)$ can be easily calculated from $\rho_\text{F}$ and $\rho_\text{B}$
by applying the temperature
derivative to the Fermi and Bose functions in Eq.~(\ref{Eq:Ener_MaxEnt}).
The results obtained in this way are in good agreement with those
from Eq.~(\ref{Eq:CV}) over a large temperature range.
However, for some parameters we observe poor convergence
especially at low temperatures because the fitting ansatz becomes too restrictive.
Therefore, we prefer the unbiased and hence superior covariance estimator for
$C_V$ and include the continuous fits from $E(T)$ merely as a guide to the eye.
As the covariance estimators for $C_V$ and $\kappa$ are subject
to large statistical fluctuations, we restrict our simulations to $L=162$
lattice sites.

To interpret the low-temperature features of the thermodynamic observables,
we also calculated the single-particle spectral functions of electrons and phonons
with Lehmann representations
\begin{align}
\nonumber%
&A(k,\omega)
	=
	\frac{1}{Z} \sum_{mn} e^{-\beta E_m} \big(1+e^{-\beta\omega}\big)
	  \absolute{\braket{m | \fan{k} | n}}^2 
	\delta(\omega - \Delta_{nm}) \, ,
\\
&B_Q(q,\omega)
	=
	\frac{M\omz^2}{Z} \sum_{mn} e^{-\beta E_m} \absolute{\braket{m | \Q{q} | n}}^2
	\delta(\omega - \Delta_{nm}) \, ,
\label{Eq:SpecB}
\end{align}
respectively.
Here, $\ket{m}$ is a many-particle eigenstate of the Hamiltonian, $E_m$ the corresponding
energy eigenvalue, and $\Delta_{nm} = E_n - E_m$.
We obtained the spectral functions from the corresponding Green's functions
$G(r,\tau) = \expvtext{\fcr{r}(\tau) \fan{0}(0)}$ and $D_Q(r,\tau) = \expvtext{\Q{r}(\tau)\Q{0}(0)}$.
The electronic Green's function can be accessed directly during the construction of
the directed loop \cite{PhysRevE.64.066701}. In the simulation of retarded interactions,
each Monte Carlo vertex already includes imaginary-time variables so that
an additional mapping is not necessary.
The phonon propagator
can be inferred from the density structure factor
$S_\rho(q,\im \Omega_m) = \int_0^\beta d\tau \, e^{\im \Omega_m \tau} \sum_r e^{-\im q r} \expvtext{\rhoan{r}(\tau) \rhoan{0}(0)}$
via~\cite{PhysRevB.91.235150}
\begin{align}
\label{Eq:Phonon_Charge}
D_Q(q,\im \Omega_m)
	=
	\Pp(\im \Omega_m)
	+ 4\lambda t \, \Pp(\im \Omega_m)^2 \, S_\rho(q,\im \Omega_m) \, .
\end{align}
Here, $\Omega_m = 2\pi m / \beta$ are the bosonic Matsubara frequencies and
$\Pp(\im\Omega_m) = \omz^2 / (\omz^2 + \Omega_m^2)$
is the free phonon propagator.
$S_\rho(q,\im\Omega_m)$ can be calculated efficiently
in the SSE representation \cite{Sandvik91,PhysRevB.56.14510}.
Finally, the spectral functions $A(k,\omega)$ and $B_Q(q,\omega)$ are obtained via
stochastic analytic continuation \cite{PhysRevB.57.10287, 2004cond.mat..3055B}
using $G(k,\tau=0)$ and $D_Q(q,\tau=0)$ as sum rules.

The total energy and hence also the specific heat $C_V = \partial E / \partial T$
are directly related to the single-particle spectral functions. Using the
equation of motion  \cite{KadanoffBaym}, we obtain the sum rule
\begin{align}
\nonumber
E
	=
	\sum_k \int_{-\infty}^\infty d\omega \, \frac{\omega + \epsilon_k}{2} \,
	n_\text{F}(\omega) \, &A(k,\omega) \\
	+ \sum_q \int_{-\infty}^\infty d\omega \, \omega \,
	n_\text{B}(\omega) \, &B(q,\omega) \, .
\label{Eq:sum_rule}
\end{align}
Here, $\epsilon_k = -2t \cos k$ is the bare electronic dispersion and
$B(q,\omega)
	=
	Z^{-1} \sum_{mn} e^{-\beta E_m} (1-e^{-\beta\omega})
	  |{\braket{m | \hat{b}_q | n}}|^2
	\delta(\omega - \Delta_{nm})$
the bosonic spectral function defined from the second-quantized operators.
In the noninteracting limit, we have $A(k,\omega) = \delta(\omega - \epsilon_k)$ and
$B(q,\omega) = [\delta(\omega - \omz) - \delta(\omega + \omz)]/2$,
\ie, the temperature dependence of $C_V$ only arises from the
distribution functions $n_\text{F}$ and $n_\text{B}$. For finite
electron-phonon interactions also
$A(k,\omega)$ and $B(q,\omega)$ change with temperature.
Note that the interaction energy equally contributes to the fermionic and
bosonic parts in Eq.~(\ref{Eq:sum_rule}). Whereas $A(k,\omega)$ has been
previously studied by an exact numerical method over the entire range of
temperatures for classical phonons \cite{PhysRevB.94.155150} [where the bosonic part in
Eq.~(\ref{Eq:sum_rule}) reduces to the classical result $L \kB T$],
the quantum case requires numerical analytic continuation and we focus on the
low-temperature spectral functions characterizing the ground state.

\section{Results}
\label{Sec:Results}

We will discuss the thermodynamic properties of the spinless Holstein model
along the paths in parameter space indicated in Fig.~\ref{Fig:PD}.
As a function of the electron-phonon interaction $\lambda$,
we consider both the antiadiabatic regime with $\omz \gg t$
and the adiabatic regime with $\omz \ll t$. Because the physics of the Holstein model
is very different in the two regimes, we calculate the
spectral functions of electrons and phonons at low temperatures to explain the characteristic
signatures that appear in the thermodynamic observables.
A special case of the adiabatic regime is the limit
$\omz = 0$ where spectral functions have been calculated exactly
at finite temperatures \cite{PhysRevB.94.155150}.
For completeness, we review the main results obtained in this limit. The effects of quantum
lattice fluctuations on the specific heat of a Peierls chain are finally studied as a function of
$\omz$ from low to high phonon frequencies.

\subsection{Polaron formation in the antiadiabatic regime}

In the antiadiabatic regime $\omega_0\gg t$, the metallic TLL phase
extends up to rather strong couplings $\lambda$, see Fig.~\ref{Fig:PD}.
With increasing $\lambda$, the electrons first undergo a crossover to small
polarons with a significantly enhanced effective mass due to the dressing
with phonons, before ordering into a polaronic superlattice at $\lambda_c$
\cite{PhysRevB.73.245120}. These effects can be characterized by the
single-particle spectral functions of electrons and phonons, which were
previously calculated numerically in the antiadiabatic regime using exact
diagonalization \cite{PhysRevB.73.245120} and a projective renormalization
approach \cite{0295-5075-76-4-644}. The electronic spectral function
has also been obtained by the bosonization method~\cite{PhysRevB.50.11179}.

\begin{figure}[tb!]
\centering
\includegraphics[width=\linewidth]{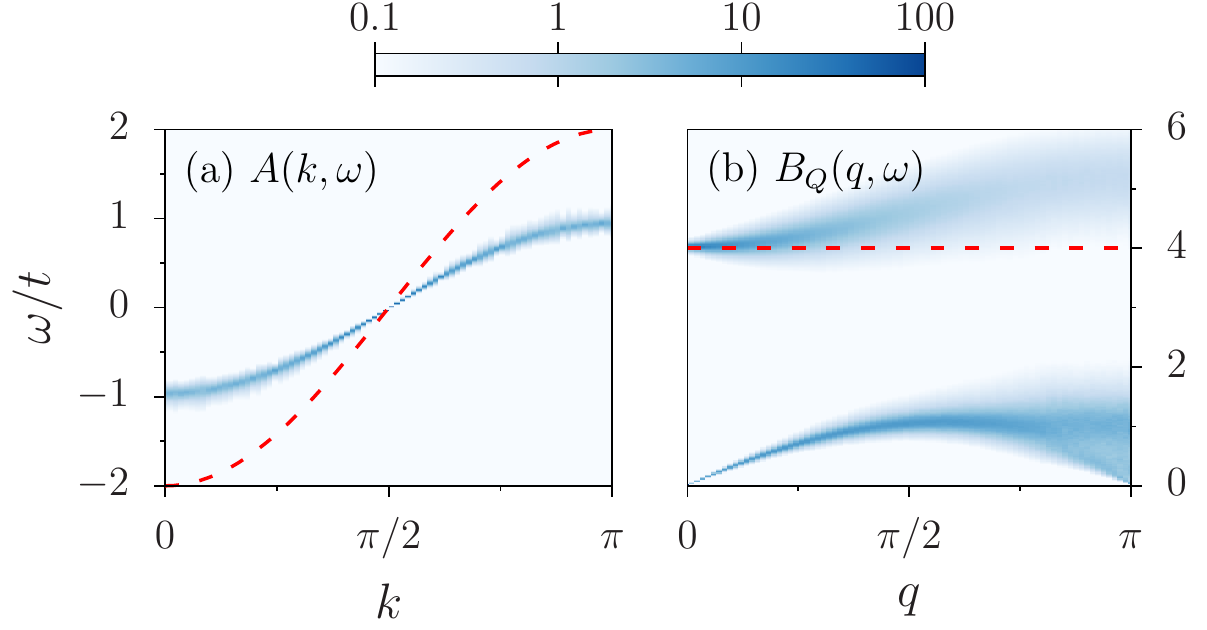}%
\caption{\label{Fig:spec_antiad}
Single-particle spectral functions of (a) electrons and (b) phonons at
$\omz/t =4$ and $\lambda=2$. Dashed lines indicate the corresponding free dispersions.
Here, $L=162$ and $\beta t = 2L$. Color scheme based on Ref.~\cite{anna_schneider_2014_10282}.
}
\end{figure}
In Fig.~\ref{Fig:spec_antiad}, we present QMC results for $A(k,\omega)$ and $B_Q(q,\omega)$
for $\omz/t=4$ and $\lambda=2$ obtained for $L=162$ and $\beta t =
2L$. The electronic spectrum in Fig.~\ref{Fig:spec_antiad}(a) exhibits a
well-defined band with a renormalized cosine dispersion $-2\tilde{t}\cos k$
and $\tilde{t}=u/2$, with  $u\approx0.47 \, \vF$ for the parameters considered.
This polaronic renormalization (but not the Peierls transition at larger
$\lambda$) can be qualitatively captured by the Lang-Firsov approximation
\cite{LangFirsov,0953-8984-18-8-011}. The renormalization of the electronic band
is also visible in the phonon spectrum in Fig.~\ref{Fig:spec_antiad}(b). The
lower branch of $B_Q(q,\omega)$ corresponds to the particle-hole continuum,
which is visible in the phonon spectral function because of the
density-displacement coupling in Eq.~(\ref{eq:holsteinmodel}) and again
reveals the renormalized electronic band. The upper branch starts at $\omega = \omz$
for $q=0$ and hardens with increasing $q$. The Peierls transition in the
antiadiabatic regime is characterized as a central-mode transition,
with a hardening of the phonon frequency and a central peak at $q=2\kF$ for
$\lambda\geq \lambda_c$ \cite{PhysRevB.73.245120}.

\begin{figure}[tb!]
\centering
\includegraphics[width=\linewidth]{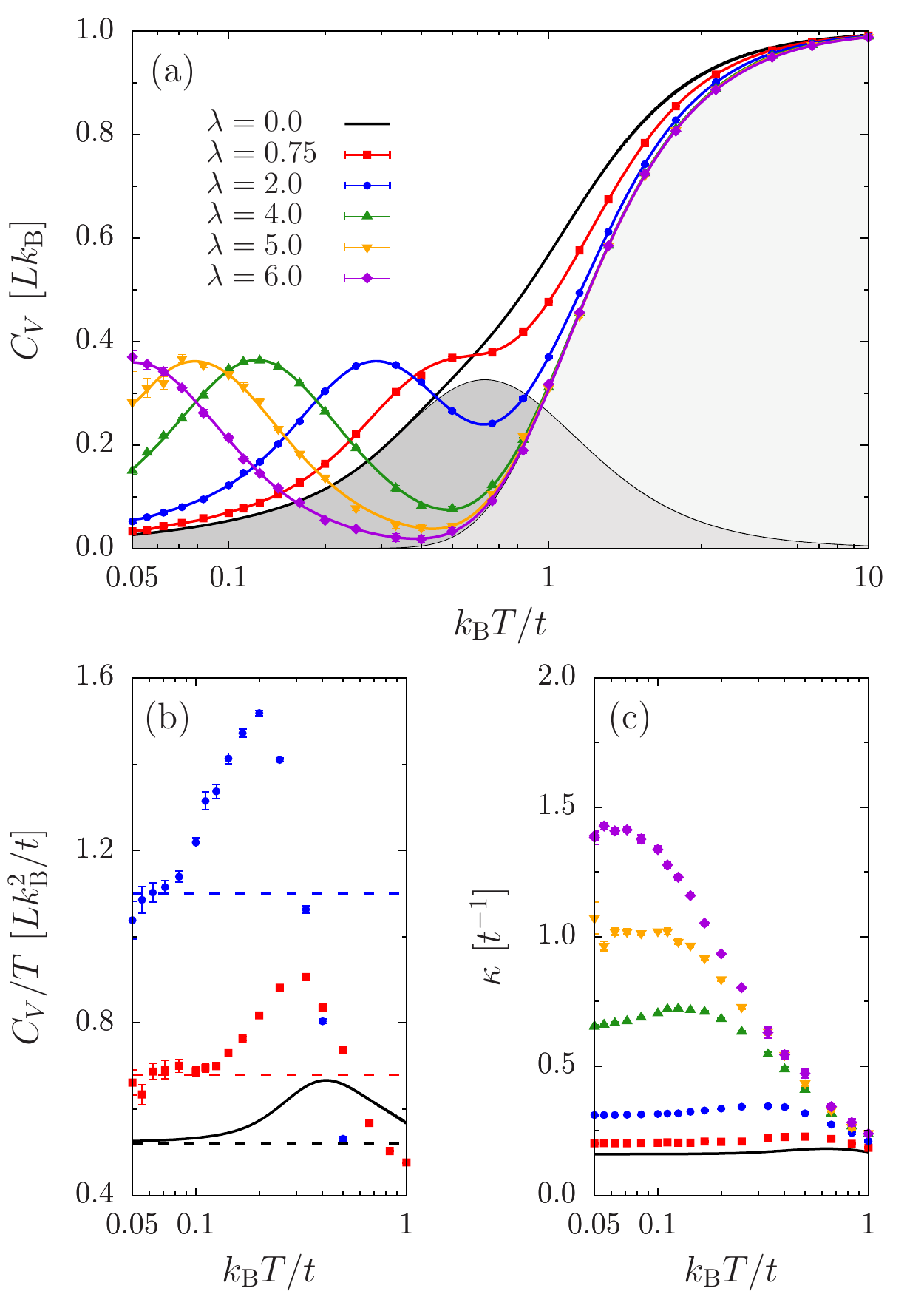}%
\caption{\label{Fig:CV_antiad}
(a) Specific heat $C_V$, (b) specific heat over temperature $C_V/T$, and
(c) compressibility $\kappa$ in the antiadiabatic regime ($\omz/t=4$, $L=162$).
Data points correspond to direct estimates, whereas straight lines in (a)
are obtained from fits to the total energy using the maximum entropy method.
The shaded areas in (a) indicate the free-electron
and free-phonon contributions to $C_V$.
The dashed lines in (b) correspond to fits to the total energy of the form
$E(T) = E_0 + \frac{1}{2} a T^2$ in the interval $T\in [0.05,0.1]$.
}
\end{figure}

To explore the thermodynamic signatures of the TLL phase, we follow the path
in Fig.~\ref{Fig:PD} at constant $\omz/t = 4$ and $L=162$.  At $\lambda=0$,
the specific heat shown in Fig.~\ref{Fig:CV_antiad}(a) is the sum of two
contributions. The free-phonon part approaches the Dulong-Petit law
$C_V = L \kB T$ for $T\to \infty$ but eventually drops off
exponentially below $\kB T \approx \omz$. The
free-electron part has a maximum at $\kB T \approx 0.63t$ that can be
identified with the onset of coherent electronic motion \cite{PhysRevB.94.155150}. 

The interpretation of the results in terms of electron and phonon
contributions remains useful at $\lambda>0$. With increasing coupling,
the high-temperature part of $C_V$ converges to the free-phonon contribution
because the renormalization of the phonon branch in $B_Q(q,\omega)$
is smeared out by thermal fluctuations. At the same time, the formation of small polarons
with substantially increased mass leads to a significant reduction of the effective
hopping $\tilde{t}$, causing the electronic contribution to $C_V$ to shift
towards lower temperatures while maintaining its shape.
In particular, the temperature dependence of $C_V$ seems to
originate mainly from the distribution functions in Eq.~(\ref{Eq:sum_rule}).
Similar to $\lambda=0$, 
the low-temperature peak in $C_V$ can be identified with the onset of
coherence, determined by the renormalized hopping $\tilde{t}$.

For the spinless model considered, there is a direct relation between the
electronic contribution to $C_V$, the density of states at the Fermi level $N(E_\text{F})$,
and the renormalized charge velocity $u$ \cite{PhysRevB.36.968,Giamarchi:743140}. 
At temperatures where the phonon contribution is frozen out, we expect
\begin{align}
\label{Eq:CV_TLL}
C_V
=
\frac{\pi^2}{3} Lk_B^2 T \, N(E_\text{F})
=
\frac{\pi}{3} \frac{Lk_B^2 T}{u} \, .
\end{align}
The first expression is generic, the second expression holds in a TLL \cite{PhysRevB.36.968}.
Figure \ref{Fig:CV_antiad}(b) shows $C_V/T$ for the two smallest $\lambda$ considered.
The convergence of $C_V/T$ for low $T$ to a constant that increases with $\lambda$
is clear evidence for the reduction of the charge velocity $u$. Low-temperature
fits of the total energy to the form $E(T) = E_0 + \frac{1}{2} a T^2$ are in good
agreement with the $C_V$ data [\cf the dashed lines in Fig.~\ref{Fig:CV_antiad}(b)].
The reduction of $u$ can also be inferred from the compressibility
shown in Fig.~\ref{Fig:CV_antiad}(c) whose low-temperature limit is given by \cite{Giamarchi:743140}
\begin{align}
\label{Eq:kappa}
\kappa
=
\frac{K}{u \pi} \, .
\end{align}
Equation (\ref{Eq:kappa}) additionally includes the Luttinger parameter $K$
that also decreases with increasing $\lambda$. Comparing
Fig.~\ref{Fig:CV_antiad}(c) with Fig.~\ref{Fig:CV_antiad}(a) reveals that the
TLL regime with a constant $\kappa$ emerges below the coherence scale
defined by the low-temperature peak in $C_V$. The observed decrease of $u$
with increasing $\lambda$ (reflecting the enhanced polaron mass) is in
contrast to the $t$-$V$ model. The latter also has a critical point
separating a TLL from a CDW insulator, but $u$ increases with increasing $V$
in the metallic phase \cite{Giamarchi:743140} as recently also observed
directly from thermodynamic properties \cite{PhysRevB.86.155156}. The
opposite behavior in electron-phonon models, namely a decrease of $u$ upon
increasing the interaction, agrees with previous numerical
\cite{PhysRevB.58.13526,PhysRevLett.80.5607} and bosonization results \cite{PhysRevB.36.968}.

\subsection{Formation of CDW order in the adiabatic limit}

The previous section revealed the principal features of $C_V$ and $\kappa$ in
the metallic phase. To understand the impact of quantum lattice fluctuations
on the thermodynamic properties of Peierls insulators, we start from the
adiabatic limit $\omz = 0$ where they are entirely absent. Then, the ground state is a Peierls
insulator for any $\lambda>0$ and exactly described by mean-field theory
\cite{Frohlich54,Peierls55}. The formation of a $2\kF$ CDW is accompanied by
the opening of a single-particle gap and the formation of shadow bands due to
the doubling of the unit cell \cite{Voit501,PhysRevB.94.155150}.

\begin{figure}[tbp]
\centering
\includegraphics[width=\linewidth]{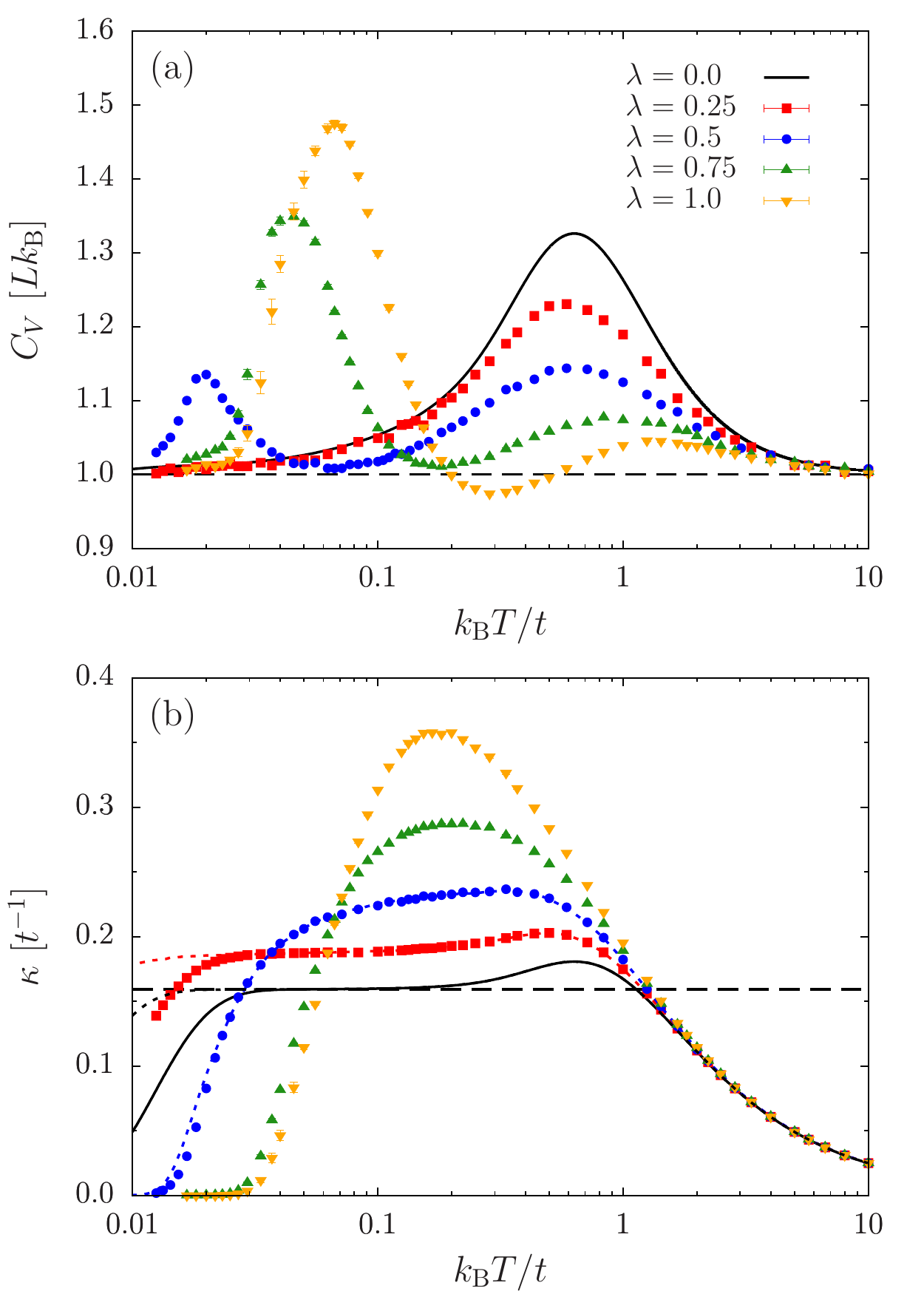}%
\caption{\label{Fig:Thermo_Classical}
(a) Specific heat $C_V$ and (b) compressibility $\kappa$ at $\omz = 0$. Dashed lines
indicate the free-phonon contribution to $C_V$ and the $T=0$ limit of $\kappa$
at $\lambda=0$. Results were obtained using the Monte Carlo method
of Ref.~\cite{1996MPLB...10..467M} for $L=162$. To illustrate finite-size
effects for $\kappa$, we also show results for $L=322$ (short-dashed lines).
}
\end{figure}

Thermal fluctuations in Peierls chains have been studied very generically
in fluctuating gap models \cite{PhysRevB.6.3409,PhysRevLett.31.462}
and recently also by QMC simulations of the classical Holstein model \cite{PhysRevB.94.155150}. The latter approach
permits the exact calculation of spectral properties without the need of numerical analytic continuation
\cite{1996MPLB...10..467M}. At $T>0$, the mean-field gap is filled in
by polaron excitations bound to thermally generated domain walls
\cite{PhysRevB.94.155150}. The temperature scale where the gap disappears 
matches the position of a low-temperature peak in the specific heat
\cite{PhysRevB.94.155150}. According to Eq.~(\ref{Eq:CV_TLL}), the
low-temperature electronic contribution to $C_V$ scales directly with the
density of states at $E_\text{F}$. Figure \ref{Fig:Thermo_Classical}(a) shows $C_V$ for different electron-phonon
couplings $\lambda$. While for $\lambda = 0.25$ the peak related to the
Peierls gap still lies outside the temperature range shown, it shifts to
higher temperatures and grows with increasing $\lambda$, in accordance with
the exponential opening of the gap. In contrast to the discontinuous feature
predicted by mean-field theory, the peak in $C_V$ is smeared out by thermal
fluctuations and appears at an energy scale much lower than the mean-field
critical temperature \cite{PhysRevLett.31.462}. At higher temperatures, $C_V$
again exhibits a peak related to the temperature scale where coherent band
motion and Fermi statistics become relevant. With increasing $\lambda$, this
peak is strongly suppressed. Whereas the Dulong-Petit law is obeyed at high
temperatures, the classical phonons produce the same constant also at $T=0$,
leading to the well-known violation of the third law of thermodynamics. 

The formation of a pseudogap at low temperatures can also be inferred from the
compressibility. Figure~\ref{Fig:Thermo_Classical}(b) reveals that $\kappa$ is suppressed at
a temperature scale that matches the peak position in $C_V$. The sharp
drop-off below $T\lesssim 0.02t$ visible in
Fig.~\ref{Fig:Thermo_Classical}(b) for $\lambda=0$ and $\lambda=0.25$
is related to a finite-size gap, as illustrated by the results for $L=322$
(short-dashed lines). If the Peierls gap is sufficiently small, $\kappa$ exhibits
the constant behavior characteristic of the TLL phase at intermediate
temperatures. Apart from that, electron-phonon coupling enhances charge fluctuations
at intermediate temperatures.

\subsection{Peierls transition in the adiabatic regime}

\begin{figure}[tbp]
\centering
\includegraphics[width=\linewidth]{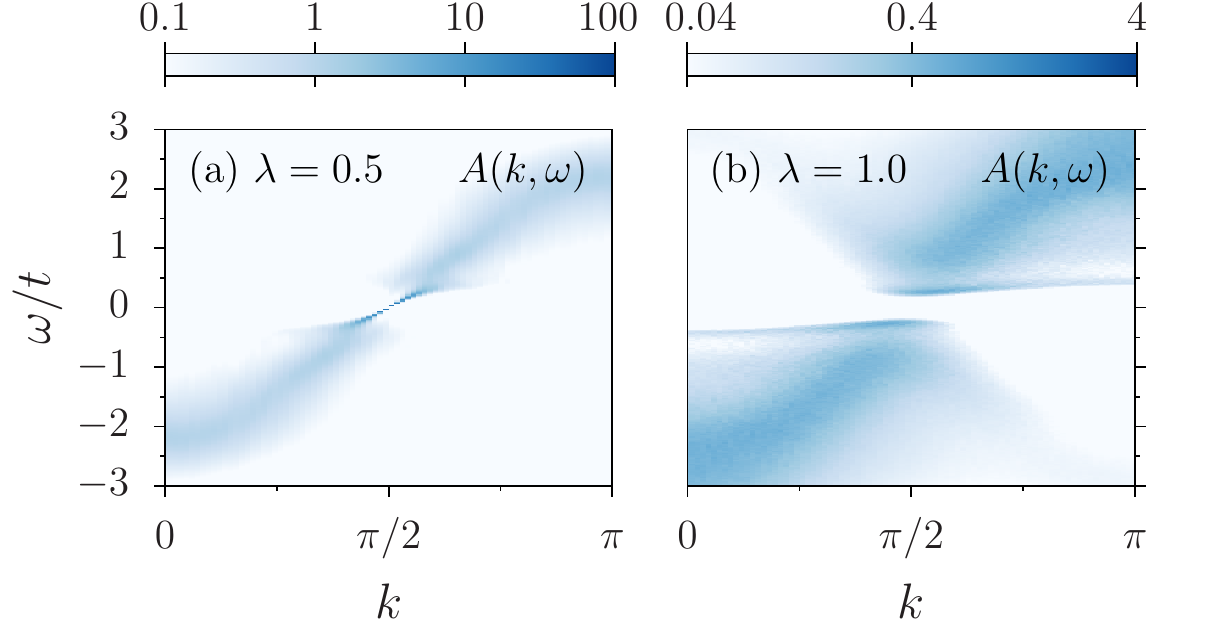}%
\caption{\label{Fig:spec_ad_el}
  Electronic single-particle spectral function $A(k,\omega)$ in (a) the
  metallic phase and (b) the Peierls phase. Close-ups of the same results are
  presented in Fig.~\ref{Fig:spec_ad}(b) and
  Fig.~\ref{Fig:spec_ad}(d), respectively. Here, $\omz/t=0.4$, $L=162$, and
  $\beta t = 2L$.
}
\end{figure}

\begin{figure}[tbp]
\centering
\includegraphics[width=\linewidth]{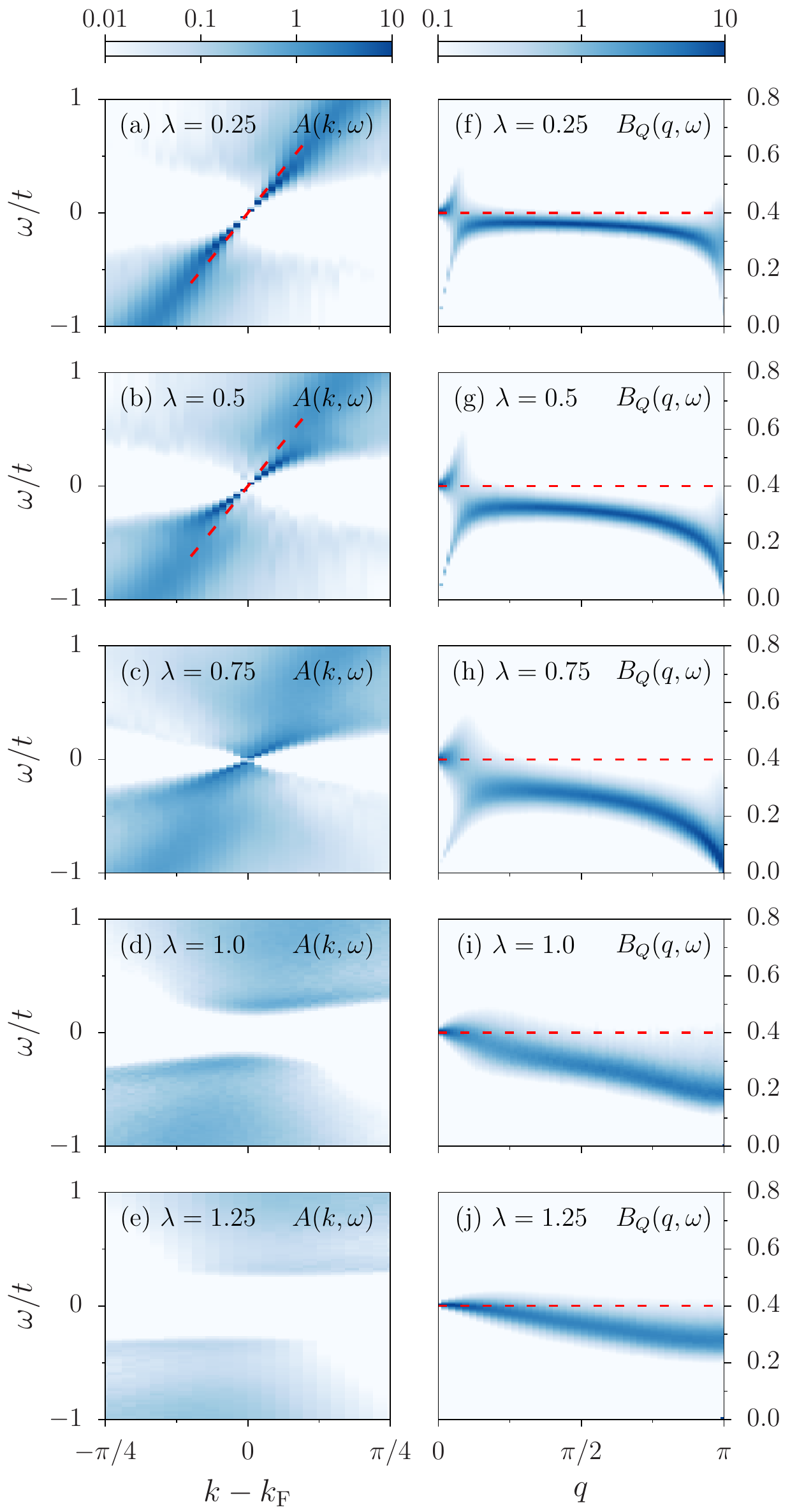}%
\caption{\label{Fig:spec_ad}
(a)--(e) Electronic single-particle spectral function $A(k,\omega)$
and (f)--(j) phonon spectral function $B_Q(q,\omega)$ for $\omz/t=0.4$,
$\beta t = 2L$, and $L=162$ (82) for $\lambda\leq 1.0$ ($\lambda=1.25$).
Dashed lines indicate the corresponding free dispersions.
}
\end{figure}

Having established the  thermodynamic signatures of
the metallic and the insulating phase, we now consider
the Peierls transition between these phases in the adiabatic quantum-phonon
regime $0<\omega_0<t$. We will see that the evolution of the low-temperature
specific heat across the transition is rather intricate due to impact of the
electron and phonon dynamics. Therefore, we first review the corresponding
single-particle spectral functions;  a more detailed discussion
can be found in Refs.~\cite{PhysRevB.83.115105,PhysRevB.91.235150}. Specifically,
Figures~\ref{Fig:spec_ad_el} and~\ref{Fig:spec_ad} show electron and phonon
spectral functions for $\omz/t = 0.4$ and different $\lambda$. The critical
coupling for the Peierls transition is $\lambda_c = 0.68(1)$
\cite{PhysRevLett.119.097401}. These results were obtained for $L= 162$ and
$\beta t = 2L$, significantly larger than in previous works. 


The electronic spectral function over the relevant energy range set by the
free bandwidth is shown in Fig.~\ref{Fig:spec_ad_el};
Figs.~\ref{Fig:spec_ad}(a)-(e) focus on the low-energy region around $E_\text{F}$.
In the TLL phase, the main effect of the electron-phonon interaction
is a renormalization of $A(k,\omega)$ inside the coherent interval
$[-\omz,\omz]$. While this effect is still small at $\lambda=0.25$
[Fig.~\ref{Fig:spec_ad}(a)], the charge velocity is significantly reduced at
$\lambda=0.5$ and a low-energy polaron band starts to split from the
incoherent high-energy excitations [Fig.~\ref{Fig:spec_ad_el}(a)]. 
The evolution of $A(k,\omega)$ in the metallic phase
can be understood in the framework of the bosonization in terms of a
hybridization of charge and phonon modes \cite{PhysRevB.50.11179}.
At $\lambda_c$, a gap opens in the polaron band which is still small at
$\lambda=0.75$ [Fig.~\ref{Fig:spec_ad}(c)] but well developed at
$\lambda=1.0$ [Fig.~\ref{Fig:spec_ad}(d)]. Finally, at $\lambda=1.25$,
the low-energy polaron excitations have almost vanished. The high-energy
features of the spectrum are dominated by mean-field-like bands, which become
more incoherent with increasing $\lambda$. At the Peierls transition, these
bands split from the polaron band and exhibit the shadow bands characteristic
for the ordered phase [Fig.~\ref{Fig:spec_ad_el}(b)]. 

The corresponding phonon spectral functions are shown in Figs.~\ref{Fig:spec_ad}(f)--(j).
In the adiabatic regime considered, the Peierls transition is a soft-mode
transition. Even at small $\lambda$, $B_Q(q,\omega)$ is significantly
renormalized near $q=\pi$ before becoming completely soft at $\lambda_c$
[Fig.~\ref{Fig:spec_ad}(h)]. In the Peierls phase, $B_Q(q,\omega)$ hardens
again and has almost returned to the original, constant dispersion for
$\lambda=1.25$. The existence of long-range order is again reflected in 
a central peak at $q=\pi$. As pointed out before, $B_Q(q,\omega)$ also contains
spectral information about the particle-hole continuum. While the high-energy part 
of $B_Q(q,\omega)$ has very small spectral weight, there is a clear feature
at small $q$ related to the hybridization of the free phonon dispersion
and the particle-hole continuum which is smeared out with increasing
$\lambda$ and disappears in the Peierls phase. Similarly, the phonon softening near $q=\pi$ can also
be regarded as a hybridization effect: Because of the presence of the
particle-hole continuum, $B_Q(q,\omega)$ must include gapless excitations at $q=\pi$ throughout the metallic phase.
Similar results for $B_Q(q,\omega)$ were previously obtained from analytic
approaches \cite{0295-5075-76-4-644} and from QMC simulations of
spin-phonon models \cite{arXiv:0705.0799, Assaad2008}.

\begin{figure}[tbp]
\centering
\includegraphics[width=\linewidth]{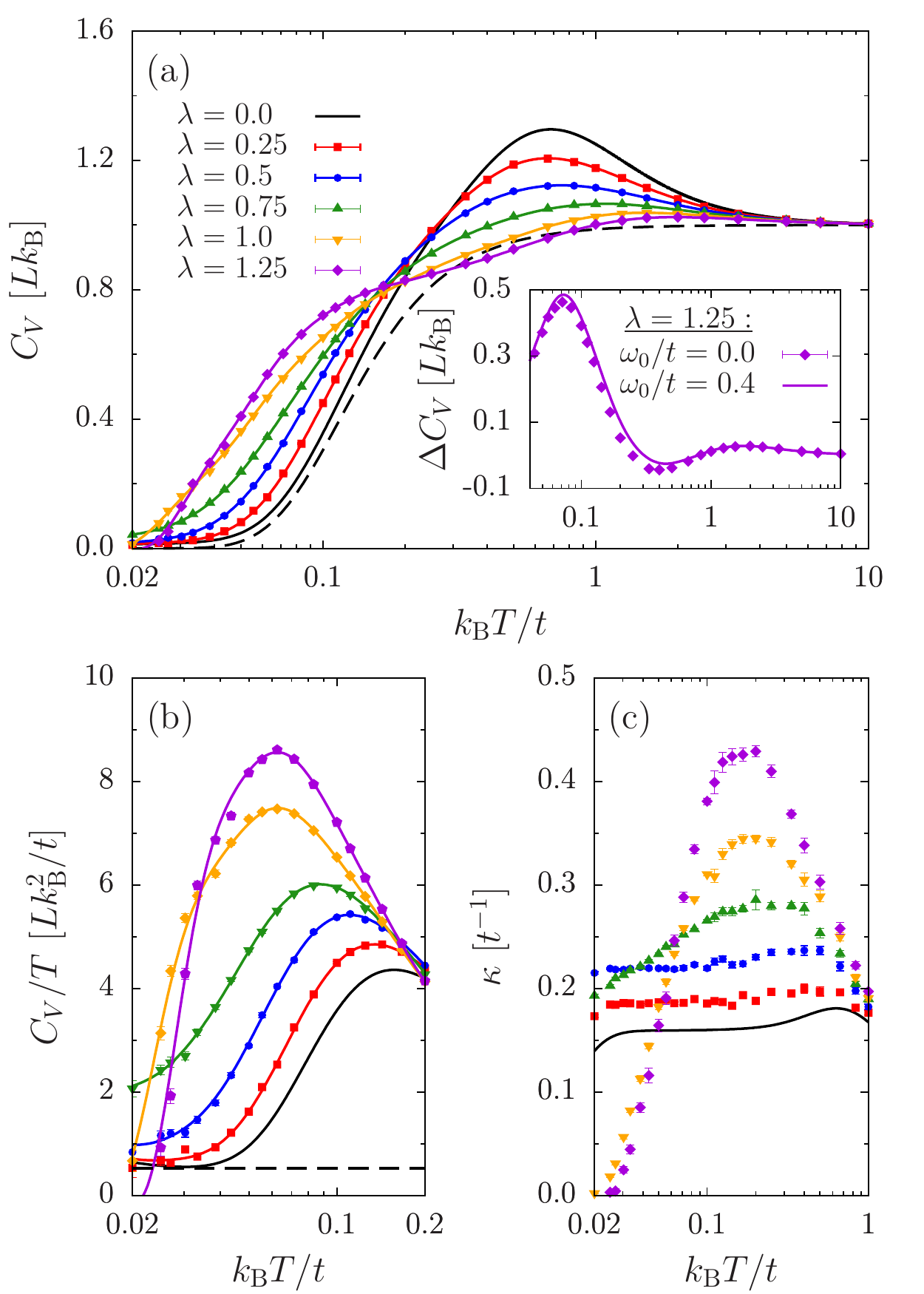}%
\caption{\label{Fig:CV_ad}
(a) Specific heat $C_V$, (b) specific heat over temperature $C_V/T$, and
(c) compressibility $\kappa$ in the adiabatic regime ($\omz/t = 0.4$, $L=162$).
Data points correspond to direct estimates, whereas straight lines in (a)
are obtained from fits to the total energy using the maximum entropy method.
The dashed line in (a) corresponds to the free-phonon contribution, whereas
the dashed line in (b) indicates the TLL result for $\lambda=0$.
The inset in (a) shows $C_V$ for $\lambda=1.25$ minus the free-phonon
contribution and compared to $\omega_0=0$.
}
\end{figure}
Figure~\ref{Fig:CV_ad} shows the evolution of $C_V$ and $\kappa$
from weak to strong coupling at $\omega_0/t=0.4$.
The specific heat in Fig.~\ref{Fig:CV_ad}(a) exhibits
a high-temperature electronic peak at $\kB T = \mathcal{O}(t)$
that is suppressed by the electron-phonon interaction, similar to $\omz=0$.
However, quantum lattice fluctuations lead to a very different behavior at
$\kB T \lesssim \omz$.
Most notably, $C_V\to 0$ for $T \to 0$ as expected
from the third law of thermodynamics. 

To better contrast the low-temperature features of the metallic and 
insulating phases, we compare $C_V/T$ in Fig.~\ref{Fig:CV_ad}(b)
to $\kappa$ in Fig.~\ref{Fig:CV_ad}(c).
In contrast to the antiadiabatic regime, the low-energy phonon mode makes
a substantial contribution to $C_V$ that only vanishes at the lowest temperatures
considered. For $\lambda=0$, we can still identify the constant contribution to $C_V/T$
expected from Eq.~(\ref{Eq:CV_TLL}) (dashed line,  $u=\vF$) at low temperatures, although
finite-size effects eventually become visible as $T\to 0$. For $\lambda>0$,
the phonon softening around $q=\pi$ enhances $C_V$ at low $T$ and thereby complicates
the analysis of the TLL behavior. From the maximum entropy fits to the total energy
(solid lines), we deduce a reduction of the charge velocity with increasing
$\lambda$ in accordance with Figs.~\ref{Fig:spec_ad}(a) and (b).
However, we cannot unambiguously determine $u$ from the QMC data
because $C_V/T$ does not yet reach a plateau for the present temperatures and
system size. By contrast, the compressibility in Fig.~\ref{Fig:CV_ad}(c)
does exhibit the expected constant behavior over a broad temperature range
before finite-size effects set in.
Even the small Peierls gap at $\lambda=0.75$ leads to a significant decrease
of $\kappa$ at $\kB T/t < 0.1$, whereas it does not leave any signature in $C_V/T$.
Deep in the Peierls phase, for $\lambda=1.0$ and $\lambda=1.25$,
$C_V/T, \kappa \to 0$
for $T \to 0$, as expected for a gapped system. 

A comparison between Fig.~\ref{Fig:CV_ad}(c) and Fig.~\ref{Fig:Thermo_Classical}(b)
reveals that the temperature dependence of $\kappa$ in the Peierls phase is
very similar to the classical case. For a direct comparison at  $\lambda=1.25$,
the inset of Fig.~\ref{Fig:CV_ad}(a) shows $\Delta C_V$, corresponding to 
$C_V$ minus the temperature-dependent free-phonon contribution. Subtracting
the free-phonon part appears justified since the phonon dispersion in
Fig.~\ref{Fig:spec_ad}(j) exhibits only minor renormalization effects
compared to the noninteracting case. The comparison reveals good agreement
between $\Delta C_V$ for $\omega_0/t=0.4$ and the adiabatic results, with
minor differences only at intermediate temperatures. This suggests that
deep in the Peierls phase the adiabatic approximation is valid
and the opening of a pseudogap occurs at the same temperature
scale as for $\omz = 0$. The same holds for the electronic spectral
function in Fig.~\ref{Fig:spec_ad}(e) which qualitatively resembles the
mean-field band structure. However, the gaps in
Fig.~\ref{Fig:spec_ad_el}(b) and Fig.~\ref{Fig:spec_ad}(e) are smaller than
$\Delta_\text{MF}$ (not shown).

\subsection{Crossover from low to high phonon frequencies}

\begin{figure}[tbp!]
\centering
\includegraphics[width=\linewidth]{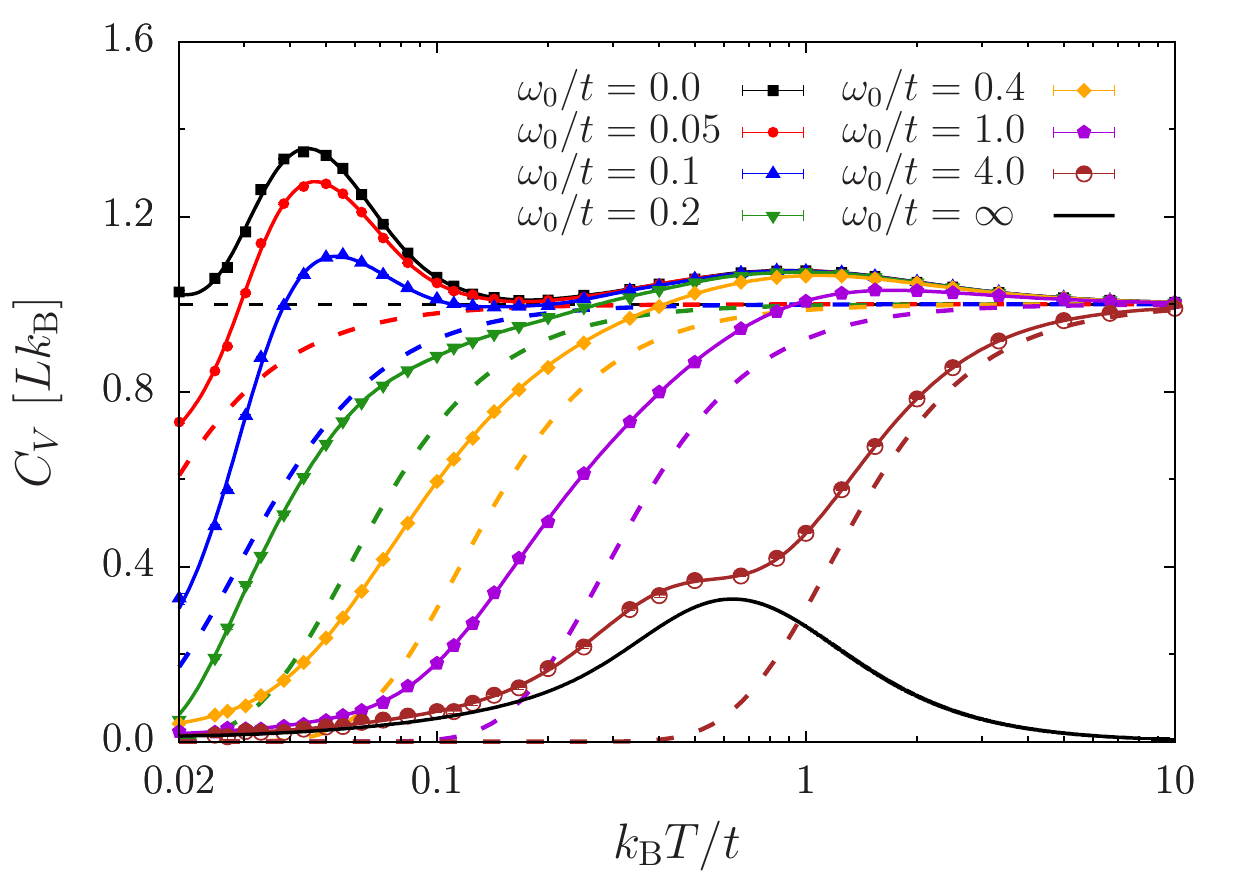}%
\caption{\label{Fig:CV_cl_to_q}
Specific heat $C_V$ for different $\omz$ at $\lambda=0.75$ and $L=162$.
Dashed lines indicate the free-phonon contributions.
}
\end{figure}

Finally, we exploit the unique advantages of our method to investigate the
impact of quantum lattice fluctuations by calculating $C_V$ over the entire
range of phonon frequencies from the adiabatic to the antiadiabatic limit at
$\lambda=0.75$. Figure~\ref{Fig:CV_cl_to_q} reveals that the adiabatic
approximation is valid for $\kB T \gtrsim \omz$. However, significant changes
are visible at lower temperatures. For $\omz/t\lesssim 0.2$, we can approximate
the phonon contribution by the noninteracting result (dashed lines), which
drops to zero at a temperature scale that increases with increasing $\omz$.
The low-temperature peak associated with pseudogap formation remains almost
unchanged for $\omz/t \leq 0.1$. It can be identified even at $\omz/t = 0.2$
after subtracting the free-phonon part (not shown). This suggests that the
coherence temperature below which the 1D Peierls physics emerges remains
almost unchanged for $\omz \ll \Delta_\text{MF}$. With increasing $\omz$, the
formation of lattice defects in the dimerization pattern is accompanied by
the creation of low-lying polaron states in $A(k,\omega)$ and a
renormalization of the phonon dispersion in $B_Q(q,\omega)$ near $q=\pi$ (see
Fig.~\ref{Fig:spec_ad}). Eventually, both excitations become gapless at the
critical value $\omega_{0,c}/t \gtrsim 0.4$ for the Peierls transition and
leave a dominant low-temperature tail in $C_V$. For $\omz \gg \Delta_\text{MF}$
we recover separate electron and phonon contributions and $C_V$ approaches
the result for noninteracting electrons as $\omz \to \infty$.

\section{Conclusions \& Outlook}
\label{Sec:Conclusion}

We studied the effects of quantum lattice fluctuations on the thermodynamic
properties of Peierls chains using the paradigmatic spinless Holstein model.
By means of a recently developed, highly efficient QMC method
\cite{PhysRevLett.119.097401}, we obtained accurate results for the specific
heat over the entire range of model parameters. These
results were complemented by calculations of the compressibility as well as
electron and phonon spectral functions.

For classical phonons, the ground state is a Peierls insulator for any coupling.
The specific heat exhibits a peak in the temperature range where dominant CDW
correlations are suppressed and the Peierls gap is filled in. A second peak
at higher temperatures is associated with the onset of coherence in the
electronic spectrum. Deep in the Peierls phase, the effects of quantum
lattice fluctuations are overall small and mostly restricted to the phonon
contribution to $C_V$. On approaching the Peierls transition in the adiabatic
regime, polaron excitations appear in the electronic spectrum and become gapless
at the critical point.  Moreover, in the adiabatic regime, the phonon spectrum exhibits
a soft mode at the transition. Both types of low-energy excitations have a
significant effect on $C_V$ at low temperatures. By contrast, in the
nonadiabatic regime, electrons are strongly renormalized by polaronic effects
even in the metallic phase. We were able to identify the expected, linear electronic
contribution to $C_V$ proportional to the charge velocity. The
renormalization of the latter was found to be particularly strong in the
antiadiabatic regime, causing a shift of the electronic contribution to $C_V$
and hence the coherence scale to lower temperatures.

Regarding the thermodynamics of 1D Peierls systems, interesting open
questions include the influence of the generic spin gap of the Luther-Emery
phase and the interplay between electron-phonon and electron-electron interactions. 
Our exact 1D results provide the starting point for a systematic
understanding of the experimental situation of quasi-1D materials in the
framework of higher-dimensional models.

\acknowledgments

Work at the University of W\"urzburg was supported by the German Research
Foundation (DFG) through SFB 1170 ToCoTronics and FOR 1807.
Work at Georgetown University was supported by the U.S. Department of
Energy (DOE), Office of Science, Basic Energy Sciences (BES) under Award
DE-FG02-08ER46542.
The authors gratefully acknowledge the computing
time granted by the John von Neumann Institute for
Computing (NIC) and provided on the supercomputer
JURECA \cite{Juelich} at the J\"ulich Supercomputing Centre.

\appendix*

\section{Direct Monte Carlo estimator for the specific heat of the Holstein model}

The SSE representation was originally formulated for instantaneous
interactions, in which case it
corresponds to a series expansion of the partition function
in the total Hamiltonian. Therefore, the specific heat has the particularly
simple estimator $C_V =\expvtext{n^2} - \expvtext{n}^2 - \expvtext{n}$,
corresponding to the fluctuations of the expansion order. To efficiently simulate fermion-boson models,
we integrate over the bosonic fields and expand in terms of retarded interactions.
As a result, we lose direct access to the bosonic fields and hence
the Hamiltonian and can no longer use the above estimator for $C_V$. We have shown in
Ref.~\cite{PhysRevB.94.245138} that the bosonic fields can be recovered from sum rules
over fermionic correlation functions using generating functionals. Moreover,
the total energy can be calculated efficiently from the distribution
of vertices using the properties of the perturbation expansion \cite{PhysRevB.56.14510}. In the following,
we show that even the second moment of the Hamiltonian can be calculated in $\mathcal{O}(n)$
operations from the distribution of vertices. To set the notation, we begin with
a brief discussion of the interaction vertex of the Holstein model.
For completeness, we first outline
the estimator for the total energy before turning to the estimator
for the second moment of the Hamiltonian.

\subsection{Interaction vertex of the Holstein model}

The directed-loop algorithm for retarded interactions is based on the generic formulation
of the perturbation expansion in the path-integral representation discussed in
Ref.~\cite{PhysRevB.94.245138}. The Monte Carlo sampling is over configurations $C=\{n,C_n,\ket{\alpha}\}$
defined by the expansion order $n$, the ordered vertex list
$C_n = \{ \nu_1, \dots, \nu_n \}$, and the state $\ket{\alpha} $ in the local occupation
number basis. 
In Ref.~\cite{PhysRevLett.119.097401}, we defined the interaction vertex for the spinless Holstein model.
In the following, we extend it to the spinful case,
where each subvertex $j\in\{1,2\}$ now has local variables $\{ a_j, b, \sigma_j, \tau_j\}$
labeling its operator type, bond, spin, and imaginary time.
The interaction vertex becomes
\begin{align}\label{eq:S1vert_spin}
\S_1 = - \iint_0^\beta d\tau_1 d\tau_2 \, \Pp(\tau_1-\tau_2)
\sum_{\substack{a_1, a_2, b, \\ \sigma_1,\sigma_2}} h_{a_1a_2,b}^{\sigma_1\sigma_2}(\tau_1,\tau_2) \, .
\end{align}
Here and in the following, we use the (anti-)symmetrized phonon propagators
$\Ppm(\tau) = \frac{1}{2}\left[ \Pnoq(\tau) \pm \Pnoq(\beta-\tau) \right]$
with
$P(\tau) = \omz \exp(-\omz \tau) / [1-\exp(-\omz\beta)]$.
The off-diagonal hopping vertices are given by 
\begin{align}
\nonumber
h_{10,b}^{\sigma_1\sigma_2}(\tau_1,\tau_2)
	&=
	\frac{t}{2 N_\sigma} \, B_{b,\sigma_1}(\tau_1) \, \mathbb{1}_{b,\sigma_2}(\tau_2) \, ,
\\
h_{01,b}^{\sigma_1\sigma_2}(\tau_1,\tau_2)
	&=
	\frac{t}{2N_\sigma} \, \mathbb{1}_{b,\sigma_1}(\tau_1) \, B_{b,\sigma_2}(\tau_2) \,,
	\label{eq:vertices_spin} 
\end{align}
whereas the diagonal interaction vertices read
\begin{align}
\nonumber
h_{22,b}^{\sigma_1\sigma_2}(\tau_1,\tau_2)
  =
    \lambda t \big[ C &+ \rhocohan{i(b),\sigma_1}(\tau_1) \, \rhocohan{i(b),\sigma_2}(\tau_2) \\
   &+ \rhocohan{j(b),\sigma_1}(\tau_1) \, \rhocohan{j(b),\sigma_2}(\tau_2) \big] 
 \label{eq:vertices2_spin} 
\end{align}
with $j(b)=i(b)+1$. We introduced an additional factor $N_\sigma$ in the hopping terms
that counts the number of spin flavors and compensates the sum over the second spin index.
For the spinful Holstein model, we have $N_\sigma = 2$, whereas the spinless case
is recovered by choosing $N_\sigma = 1$ and dropping the spin indices.
The constant shift $C$ in Eq.~(\ref{eq:vertices2_spin}) ensures positive Monte Carlo weights.
While we only consider the half-filled Holstein model, a chemical potential can be
easily included in the diagonal term. In the following, we partition the total expansion order $n = n_1 + n_2$ into
the number of off-diagonal vertices $n_1 = n_{10}+n_{01}$ and the number of diagonal
vertices $n_2 =n_{22}$.

\subsection{Total energy}

For completeness, we  review the estimator for the total energy
derived in Ref.~\cite{PhysRevB.94.245138}. The Hamiltonian of the Holstein model,
$\hat{H} = \sum_x \hat{H}_x$, is split into three contributions labeled by 
the indices $x\in\{\text{el},\text{ph},\text{ep}\}$.
The first element corresponds to the kinetic energy of the electrons,
the second to the purely bosonic part (including a shift of $\omz/2$ per site),
and the third to the electron-phonon interaction---see
Eq.~(\ref{eq:holsteinmodel}) for exact definitions.
For each Monte Carlo configuration, we define the contributions
to the total energy by
\begin{align}
E_x(C_n)=\frac{1}{\beta} \int_0^\beta d\tau \expvcn{H_x(\tau)} \, .
\end{align}
Translational invariance of all vertices is taken into account by the average over imaginary time.
Using the sum rules specified in Ref.~\cite{PhysRevB.94.245138}, each contribution to $E(C_n)$ can be expressed
in terms of the interaction vertices (\ref{eq:vertices_spin}) and
(\ref{eq:vertices2_spin}) to obtain
\begin{align}
\Eel(C_n)
	& =
	-\frac{n_1}{\beta} \, ,
	\\ \nonumber
  \Eph(C_n)
    & =
      L \, \Pp(0) - {\lambda t C L N_{\sigma}^2 } \\
      & \phantom{ = } + \sum_{k=1}^{n_2}
      \left[\Ppbar(\tau_k - \tau'_k)  - \Pmbar(\tau_k - \tau'_k) \right]
        \, ,
	\\
  \Eeph(C_n)
   & =
      - \frac{2 n_2}{\beta} + 2 \lambda t C L N_{\sigma}^2 \, .
\end{align}
Translational invariance of all vertices is contained in the averaged propagator
\begin{align}
\Ppmbar(\tau_k - \tau'_k)
	=
	\frac{1}{\beta} \int_0^{\beta}  d\tau \,
  \frac{
  \Ppm(\tau_k +\tau) \Ppm(\tau'_k +\tau)
    }{
  \Pp(\tau_k - \tau'_k)
  } \, .
 \end{align}
 Explicitly, it is given by ($\tau \in [-\beta,\beta]$)
 \begin{align}
 \nonumber
  \Ppmbar(\tau)
  = 
  \frac{1}{2\beta}
&\pm \frac{\omega_0}{4}   \frac{\beta -\absolute{\tau}}{\beta}
\left[\coth(\omega_0\beta/2) - \frac{\Pm(\tau)}{\Pp(\tau)}\right]  \\
&\pm \frac{\omega_0}{4}  \frac{\absolute{\tau}}{\beta}
\left[\coth(\omega_0\beta/2) +\frac{\Pm(\tau)}{\Pp(\tau)}\right] \, .
\end{align}

\subsection{Second moment of the Hamiltonian}

To calculate the second moment of $\hat{H}$, we write its expectation value
in a translationally invariant form, \ie,
\begin{align}
\expv{\ham{}^2}
	=
	\frac{1}{\beta^2} \iint_0^\beta d\tau d\tau' \expv{H(\tau) \, H(\tau')} \, .
\end{align}
Using the time-displaced form of the correlation function ensures that
in the end each operator identified with a subvertex of the interaction vertex obtains
an individual time label that is integrated over. We again split the total
Hamiltonian into fermionic, bosonic, and fermion-boson contributions.
To simplify the notation, we define ($x,x' \in \{ \text{el}, \text{ph}, \text{ep} \}$)
\begin{align}
F_{x-x'}(C_n) = \frac{1}{\beta^2} \iint_0^\beta d\tau d\tau' \expvcn{H_x(\tau) \, H_{x'}(\tau')} \, .
\end{align}

The estimator for the purely electronic contribution has the same form as usual
and is given by
\begin{align}
F_{\text{el}-\text{el}}(C_n)
	=
	\frac{n_1 \left( n_1 - 1\right)}{\beta^2} \, .
\end{align}
Also the mixing terms between the electronic part of the Hamiltonian and the remaining parts have
simple estimators that are given by
\begin{align}
\label{eq:mix1}
 F_{\text{el}-\text{ph}} (C_n)
	&=
	\Eel(C_n) \, \Eph(C_n)  \, , \\
 F_{\text{el}-\text{ep}} (C_n)
	&=
	 \Eel(C_n) \, \Eeph(C_n) \, .
\label{eq:mix2}
\end{align}
The electronic and the bosonic contributions are recovered from vertices
with different operator types and hence do not interfere in the total estimators. 

The derivation of estimators is more complicated for correlation functions, where each
part of the Hamiltonian contains bosonic fields. When we calculate
the functional derivatives to obtain sum rules for the bosonic fields, we have
to account for additional cross terms that do not appear for the individual energies.
For example, the correlation function between the electron-phonon parts of the Hamiltonian
becomes
\begin{align}
\nonumber
\expv{H_{\text{ep}}(\tau) \, H_{\text{ep}}(\tau')}
	= 
	4&\lambda t \, \Pp(\tau-\tau') \sum_i \expv{\rhocohan{i}(\tau)\rhocohan{i}(\tau')} \\ \nonumber
	+ (4\lambda t)^2 \iint_0^\beta d\tau_1 &d\tau_2 \, \Pp(\tau-\tau_1) \, \Pp(\tau'-\tau_2)  \\
	\times \sum_{ij} &\expv{\rhocohan{i}(\tau)\rhocohan{i}(\tau_1) \rhocohan{j}(\tau')\rhocohan{j}(\tau_2)} \, .
\end{align}
The first term on the r.h.s.~is an additional cross term.
The corresponding estimator is
\begin{align}
F_{\text{ep}-\text{ep}}(C_n)
	=
	\Eeph(C_n)^2 - \frac{4n_2}{\beta^2} - \frac{\Eeph(C_n)}{\beta} \, .
\end{align}
Similar considerations yield the estimators
\begin{align}
F_{\text{ep}-\text{ph}}(C_n)
	=
	\Eeph(C_n) \, \Eph(C_n) + \frac{2\lambda t CLN_\sigma^2}{\beta}
\end{align}
and
\begin{align}
\nonumber
F_{\text{ph}-\text{ph}}(C_n)
	&=
	 \Eph(C_n)^2 + L \, \Pp(0) \left[ \Ppbar(0) - \Pmbar(0) \right] \\ \nonumber
	&\phantom{= \ } - \sum_{k=1}^{n_2} \left[\Ppbar(\tau_k - \tau'_k)  - \Pmbar(\tau_k - \tau'_k) \right]^2 \\
	&\phantom{= \ } + \sum_{k=1}^{n_2} \frac{Z(\tau_k-\tau'_k)}{\Pp(\tau_k-\tau'_k)}
	- \frac{2\lambda t CLN_\sigma^2}{\beta} \, .
\end{align}
For the latter, we introduced an additional function
\begin{align}
\nonumber
Z(\tau)
	=
	\frac{\omz^3}{\beta^2} \, e^{ (\beta - \tau) \omz} \, n_\text{B}(\omz) \, 
	 \big[
	&\tau^2 + \beta \left(\beta + 2 \tau\right) n_\text{B}(\omz) \\
	&+ 2 \beta^2 n_\text{B}(\omz)^2
	\big]
\end{align}
that is defined for $\tau \in [0,\beta)$. To evaluate $Z(\tau)$ for $\tau<0$,
we use $Z(\tau+\beta) = Z(\tau)$. Here,
$n_\text{B}(\omega) =  [\exp(\beta \omega)-1]^{-1}$.


\begin{thebibliography}{77}%
\makeatletter
\providecommand \@ifxundefined [1]{%
 \@ifx{#1\undefined}
}%
\providecommand \@ifnum [1]{%
 \ifnum #1\expandafter \@firstoftwo
 \else \expandafter \@secondoftwo
 \fi
}%
\providecommand \@ifx [1]{%
 \ifx #1\expandafter \@firstoftwo
 \else \expandafter \@secondoftwo
 \fi
}%
\providecommand \natexlab [1]{#1}%
\providecommand \enquote  [1]{``#1''}%
\providecommand \bibnamefont  [1]{#1}%
\providecommand \bibfnamefont [1]{#1}%
\providecommand \citenamefont [1]{#1}%
\providecommand \href@noop [0]{\@secondoftwo}%
\providecommand \href [0]{\begingroup \@sanitize@url \@href}%
\providecommand \@href[1]{\@@startlink{#1}\@@href}%
\providecommand \@@href[1]{\endgroup#1\@@endlink}%
\providecommand \@sanitize@url [0]{\catcode `\\12\catcode `\$12\catcode
  `\&12\catcode `\#12\catcode `\^12\catcode `\_12\catcode `\%12\relax}%
\providecommand \@@startlink[1]{}%
\providecommand \@@endlink[0]{}%
\providecommand \url  [0]{\begingroup\@sanitize@url \@url }%
\providecommand \@url [1]{\endgroup\@href {#1}{\urlprefix }}%
\providecommand \urlprefix  [0]{URL }%
\providecommand \Eprint [0]{\href }%
\providecommand \doibase [0]{http://dx.doi.org/}%
\providecommand \selectlanguage [0]{\@gobble}%
\providecommand \bibinfo  [0]{\@secondoftwo}%
\providecommand \bibfield  [0]{\@secondoftwo}%
\providecommand \translation [1]{[#1]}%
\providecommand \BibitemOpen [0]{}%
\providecommand \bibitemStop [0]{}%
\providecommand \bibitemNoStop [0]{.\EOS\space}%
\providecommand \EOS [0]{\spacefactor3000\relax}%
\providecommand \BibitemShut  [1]{\csname bibitem#1\endcsname}%
\let\auto@bib@innerbib\@empty
\bibitem [{\citenamefont {Fr\"ohlich}(1954)}]{Frohlich54}%
  \BibitemOpen
  \bibfield  {author} {\bibinfo {author} {\bibfnamefont {H.}~\bibnamefont
  {Fr\"ohlich}},\ }\href@noop {} {\bibfield  {journal} {\bibinfo  {journal}
  {Proc. Roy. Soc. A}\ }\textbf {\bibinfo {volume} {223}},\ \bibinfo {pages}
  {296} (\bibinfo {year} {1954})}\BibitemShut {NoStop}%
\bibitem [{\citenamefont {Peierls}(1955)}]{Peierls55}%
  \BibitemOpen
  \bibfield  {author} {\bibinfo {author} {\bibfnamefont {R.~E.}\ \bibnamefont
  {Peierls}},\ }\href@noop {} {\emph {\bibinfo {title} {Quantum Theory of
  Solids}}}\ (\bibinfo  {publisher} {Clarendon Press},\ \bibinfo {address}
  {Oxford},\ \bibinfo {year} {1955})\BibitemShut {NoStop}%
\bibitem [{\citenamefont {Claessen}\ \emph {et~al.}(2002)\citenamefont
  {Claessen}, \citenamefont {Sing}, \citenamefont {Schwingenschl\"ogl},
  \citenamefont {Blaha}, \citenamefont {Dressel},\ and\ \citenamefont
  {Jacobsen}}]{PhysRevLett.88.096402}%
  \BibitemOpen
  \bibfield  {author} {\bibinfo {author} {\bibfnamefont {R.}~\bibnamefont
  {Claessen}}, \bibinfo {author} {\bibfnamefont {M.}~\bibnamefont {Sing}},
  \bibinfo {author} {\bibfnamefont {U.}~\bibnamefont {Schwingenschl\"ogl}},
  \bibinfo {author} {\bibfnamefont {P.}~\bibnamefont {Blaha}}, \bibinfo
  {author} {\bibfnamefont {M.}~\bibnamefont {Dressel}}, \ and\ \bibinfo
  {author} {\bibfnamefont {C.~S.}\ \bibnamefont {Jacobsen}},\ }\href {\doibase
  10.1103/PhysRevLett.88.096402} {\bibfield  {journal} {\bibinfo  {journal}
  {Phys. Rev. Lett.}\ }\textbf {\bibinfo {volume} {88}},\ \bibinfo {pages}
  {096402} (\bibinfo {year} {2002})}\BibitemShut {NoStop}%
\bibitem [{\citenamefont {Travaglini}\ \emph {et~al.}(1983)\citenamefont
  {Travaglini}, \citenamefont {M\"orke},\ and\ \citenamefont
  {Wachter}}]{TRAVAGLINI1983289}%
  \BibitemOpen
  \bibfield  {author} {\bibinfo {author} {\bibfnamefont {G.}~\bibnamefont
  {Travaglini}}, \bibinfo {author} {\bibfnamefont {I.}~\bibnamefont {M\"orke}},
  \ and\ \bibinfo {author} {\bibfnamefont {P.}~\bibnamefont {Wachter}},\ }\href
  {\doibase http://dx.doi.org/10.1016/0038-1098(83)90483-0} {\bibfield
  {journal} {\bibinfo  {journal} {Solid State Commun.}\ }\textbf {\bibinfo
  {volume} {45}},\ \bibinfo {pages} {289 } (\bibinfo {year}
  {1983})}\BibitemShut {NoStop}%
\bibitem [{\citenamefont {Pytte}(1974)}]{PhysRevB.10.4637}%
  \BibitemOpen
  \bibfield  {author} {\bibinfo {author} {\bibfnamefont {E.}~\bibnamefont
  {Pytte}},\ }\href {\doibase 10.1103/PhysRevB.10.4637} {\bibfield  {journal}
  {\bibinfo  {journal} {Phys. Rev. B}\ }\textbf {\bibinfo {volume} {10}},\
  \bibinfo {pages} {4637} (\bibinfo {year} {1974})}\BibitemShut {NoStop}%
\bibitem [{\citenamefont {Hase}\ \emph {et~al.}(1993)\citenamefont {Hase},
  \citenamefont {Terasaki},\ and\ \citenamefont
  {Uchinokura}}]{PhysRevLett.70.3651}%
  \BibitemOpen
  \bibfield  {author} {\bibinfo {author} {\bibfnamefont {M.}~\bibnamefont
  {Hase}}, \bibinfo {author} {\bibfnamefont {I.}~\bibnamefont {Terasaki}}, \
  and\ \bibinfo {author} {\bibfnamefont {K.}~\bibnamefont {Uchinokura}},\
  }\href {\doibase 10.1103/PhysRevLett.70.3651} {\bibfield  {journal} {\bibinfo
   {journal} {Phys. Rev. Lett.}\ }\textbf {\bibinfo {volume} {70}},\ \bibinfo
  {pages} {3651} (\bibinfo {year} {1993})}\BibitemShut {NoStop}%
\bibitem [{\citenamefont {Craven}\ \emph {et~al.}(1974)\citenamefont {Craven},
  \citenamefont {Salamon}, \citenamefont {DePasquali}, \citenamefont {Herman},
  \citenamefont {Stucky},\ and\ \citenamefont {Schultz}}]{PhysRevLett.32.769}%
  \BibitemOpen
  \bibfield  {author} {\bibinfo {author} {\bibfnamefont {R.~A.}\ \bibnamefont
  {Craven}}, \bibinfo {author} {\bibfnamefont {M.~B.}\ \bibnamefont {Salamon}},
  \bibinfo {author} {\bibfnamefont {G.}~\bibnamefont {DePasquali}}, \bibinfo
  {author} {\bibfnamefont {R.~M.}\ \bibnamefont {Herman}}, \bibinfo {author}
  {\bibfnamefont {G.}~\bibnamefont {Stucky}}, \ and\ \bibinfo {author}
  {\bibfnamefont {A.}~\bibnamefont {Schultz}},\ }\href {\doibase
  10.1103/PhysRevLett.32.769} {\bibfield  {journal} {\bibinfo  {journal} {Phys.
  Rev. Lett.}\ }\textbf {\bibinfo {volume} {32}},\ \bibinfo {pages} {769}
  (\bibinfo {year} {1974})}\BibitemShut {NoStop}%
\bibitem [{\citenamefont {Wei}\ \emph {et~al.}(1977)\citenamefont {Wei},
  \citenamefont {Heeger}, \citenamefont {Salamon},\ and\ \citenamefont
  {Delker}}]{WEI1977595}%
  \BibitemOpen
  \bibfield  {author} {\bibinfo {author} {\bibfnamefont {T.}~\bibnamefont
  {Wei}}, \bibinfo {author} {\bibfnamefont {A.~J.}\ \bibnamefont {Heeger}},
  \bibinfo {author} {\bibfnamefont {M.~B.}\ \bibnamefont {Salamon}}, \ and\
  \bibinfo {author} {\bibfnamefont {G.~E.}\ \bibnamefont {Delker}},\ }\href
  {\doibase http://dx.doi.org/10.1016/0038-1098(77)90041-2} {\bibfield
  {journal} {\bibinfo  {journal} {Solid State Commun.}\ }\textbf {\bibinfo
  {volume} {21}},\ \bibinfo {pages} {595 } (\bibinfo {year}
  {1977})}\BibitemShut {NoStop}%
\bibitem [{\citenamefont {Biljakovic}\ \emph {et~al.}(1986)\citenamefont
  {Biljakovic}, \citenamefont {Lasjaunias}, \citenamefont {Zougmore},
  \citenamefont {Monceau}, \citenamefont {Levy}, \citenamefont {Bernard},\ and\
  \citenamefont {Currat}}]{PhysRevLett.57.1907}%
  \BibitemOpen
  \bibfield  {author} {\bibinfo {author} {\bibfnamefont {K.}~\bibnamefont
  {Biljakovic}}, \bibinfo {author} {\bibfnamefont {J.~C.}\ \bibnamefont
  {Lasjaunias}}, \bibinfo {author} {\bibfnamefont {F.}~\bibnamefont
  {Zougmore}}, \bibinfo {author} {\bibfnamefont {P.}~\bibnamefont {Monceau}},
  \bibinfo {author} {\bibfnamefont {F.}~\bibnamefont {Levy}}, \bibinfo {author}
  {\bibfnamefont {L.}~\bibnamefont {Bernard}}, \ and\ \bibinfo {author}
  {\bibfnamefont {R.}~\bibnamefont {Currat}},\ }\href {\doibase
  10.1103/PhysRevLett.57.1907} {\bibfield  {journal} {\bibinfo  {journal}
  {Phys. Rev. Lett.}\ }\textbf {\bibinfo {volume} {57}},\ \bibinfo {pages}
  {1907} (\bibinfo {year} {1986})}\BibitemShut {NoStop}%
\bibitem [{\citenamefont {Liu}\ \emph {et~al.}(1995)\citenamefont {Liu},
  \citenamefont {Wosnitza}, \citenamefont {{von L\"ohneysen}},\ and\
  \citenamefont {Kremer}}]{PhysRevLett.75.771}%
  \BibitemOpen
  \bibfield  {author} {\bibinfo {author} {\bibfnamefont {X.}~\bibnamefont
  {Liu}}, \bibinfo {author} {\bibfnamefont {J.}~\bibnamefont {Wosnitza}},
  \bibinfo {author} {\bibfnamefont {H.}~\bibnamefont {{von L\"ohneysen}}}, \
  and\ \bibinfo {author} {\bibfnamefont {R.~K.}\ \bibnamefont {Kremer}},\
  }\href {\doibase 10.1103/PhysRevLett.75.771} {\bibfield  {journal} {\bibinfo
  {journal} {Phys. Rev. Lett.}\ }\textbf {\bibinfo {volume} {75}},\ \bibinfo
  {pages} {771} (\bibinfo {year} {1995})}\BibitemShut {NoStop}%
\bibitem [{\citenamefont {Powell}\ \emph {et~al.}(1998)\citenamefont {Powell},
  \citenamefont {Brill}, \citenamefont {Zeng},\ and\ \citenamefont
  {Greenblatt}}]{PhysRevB.58.R2937}%
  \BibitemOpen
  \bibfield  {author} {\bibinfo {author} {\bibfnamefont {D.~K.}\ \bibnamefont
  {Powell}}, \bibinfo {author} {\bibfnamefont {J.~W.}\ \bibnamefont {Brill}},
  \bibinfo {author} {\bibfnamefont {Z.}~\bibnamefont {Zeng}}, \ and\ \bibinfo
  {author} {\bibfnamefont {M.}~\bibnamefont {Greenblatt}},\ }\href {\doibase
  10.1103/PhysRevB.58.R2937} {\bibfield  {journal} {\bibinfo  {journal} {Phys.
  Rev. B}\ }\textbf {\bibinfo {volume} {58}},\ \bibinfo {pages} {R2937}
  (\bibinfo {year} {1998})}\BibitemShut {NoStop}%
\bibitem [{\citenamefont {Kwok}\ \emph {et~al.}(1990)\citenamefont {Kwok},
  \citenamefont {Gruner},\ and\ \citenamefont {Brown}}]{PhysRevLett.65.365}%
  \BibitemOpen
  \bibfield  {author} {\bibinfo {author} {\bibfnamefont {R.~S.}\ \bibnamefont
  {Kwok}}, \bibinfo {author} {\bibfnamefont {G.}~\bibnamefont {Gruner}}, \ and\
  \bibinfo {author} {\bibfnamefont {S.~E.}\ \bibnamefont {Brown}},\ }\href
  {\doibase 10.1103/PhysRevLett.65.365} {\bibfield  {journal} {\bibinfo
  {journal} {Phys. Rev. Lett.}\ }\textbf {\bibinfo {volume} {65}},\ \bibinfo
  {pages} {365} (\bibinfo {year} {1990})}\BibitemShut {NoStop}%
\bibitem [{\citenamefont {Pouget}(2016)}]{Pouget2016332}%
  \BibitemOpen
  \bibfield  {author} {\bibinfo {author} {\bibfnamefont {J.-P.}\ \bibnamefont
  {Pouget}},\ }\href@noop {} {\bibfield  {journal} {\bibinfo  {journal}
  {Comptes Rendus Physique}\ }\textbf {\bibinfo {volume} {17}},\ \bibinfo
  {pages} {332 } (\bibinfo {year} {2016})}\BibitemShut {NoStop}%
\bibitem [{\citenamefont {Ashcroft}\ and\ \citenamefont
  {Mermin}(1976)}]{AshcroftMermin}%
  \BibitemOpen
  \bibfield  {author} {\bibinfo {author} {\bibfnamefont {N.~W.}\ \bibnamefont
  {Ashcroft}}\ and\ \bibinfo {author} {\bibfnamefont {N.~D.}\ \bibnamefont
  {Mermin}},\ }\href@noop {} {\emph {\bibinfo {title} {Solid State Physics}}}\
  (\bibinfo  {publisher} {Saunders College Publishing},\ \bibinfo {address}
  {Philadelphia},\ \bibinfo {year} {1976})\BibitemShut {NoStop}%
\bibitem [{\citenamefont {Kuiper}(1955)}]{Kuiper55}%
  \BibitemOpen
  \bibfield  {author} {\bibinfo {author} {\bibfnamefont {C.~G.}\ \bibnamefont
  {Kuiper}},\ }\href@noop {} {\bibfield  {journal} {\bibinfo  {journal} {Proc.
  Roy. Soc. A}\ }\textbf {\bibinfo {volume} {227}},\ \bibinfo {pages} {214}
  (\bibinfo {year} {1955})}\BibitemShut {NoStop}%
\bibitem [{\citenamefont {Heeger}\ \emph {et~al.}(1988)\citenamefont {Heeger},
  \citenamefont {Kivelson}, \citenamefont {Schrieffer},\ and\ \citenamefont
  {Su}}]{RevModPhys.60.781}%
  \BibitemOpen
  \bibfield  {author} {\bibinfo {author} {\bibfnamefont {A.~J.}\ \bibnamefont
  {Heeger}}, \bibinfo {author} {\bibfnamefont {S.}~\bibnamefont {Kivelson}},
  \bibinfo {author} {\bibfnamefont {J.~R.}\ \bibnamefont {Schrieffer}}, \ and\
  \bibinfo {author} {\bibfnamefont {W.~P.}\ \bibnamefont {Su}},\ }\href
  {\doibase 10.1103/RevModPhys.60.781} {\bibfield  {journal} {\bibinfo
  {journal} {Rev. Mod. Phys.}\ }\textbf {\bibinfo {volume} {60}},\ \bibinfo
  {pages} {781} (\bibinfo {year} {1988})}\BibitemShut {NoStop}%
\bibitem [{\citenamefont {Scalapino}\ \emph {et~al.}(1972)\citenamefont
  {Scalapino}, \citenamefont {Sears},\ and\ \citenamefont
  {Ferrell}}]{PhysRevB.6.3409}%
  \BibitemOpen
  \bibfield  {author} {\bibinfo {author} {\bibfnamefont {D.~J.}\ \bibnamefont
  {Scalapino}}, \bibinfo {author} {\bibfnamefont {M.}~\bibnamefont {Sears}}, \
  and\ \bibinfo {author} {\bibfnamefont {R.~A.}\ \bibnamefont {Ferrell}},\
  }\href {\doibase 10.1103/PhysRevB.6.3409} {\bibfield  {journal} {\bibinfo
  {journal} {Phys. Rev. B}\ }\textbf {\bibinfo {volume} {6}},\ \bibinfo {pages}
  {3409} (\bibinfo {year} {1972})}\BibitemShut {NoStop}%
\bibitem [{\citenamefont {Lee}\ \emph {et~al.}(1973)\citenamefont {Lee},
  \citenamefont {Rice},\ and\ \citenamefont {Anderson}}]{PhysRevLett.31.462}%
  \BibitemOpen
  \bibfield  {author} {\bibinfo {author} {\bibfnamefont {P.~A.}\ \bibnamefont
  {Lee}}, \bibinfo {author} {\bibfnamefont {T.~M.}\ \bibnamefont {Rice}}, \
  and\ \bibinfo {author} {\bibfnamefont {P.~W.}\ \bibnamefont {Anderson}},\
  }\href {\doibase 10.1103/PhysRevLett.31.462} {\bibfield  {journal} {\bibinfo
  {journal} {Phys. Rev. Lett.}\ }\textbf {\bibinfo {volume} {31}},\ \bibinfo
  {pages} {462} (\bibinfo {year} {1973})}\BibitemShut {NoStop}%
\bibitem [{\citenamefont {Weber}\ \emph
  {et~al.}(2016{\natexlab{a}})\citenamefont {Weber}, \citenamefont {Assaad},\
  and\ \citenamefont {Hohenadler}}]{PhysRevB.94.155150}%
  \BibitemOpen
  \bibfield  {author} {\bibinfo {author} {\bibfnamefont {M.}~\bibnamefont
  {Weber}}, \bibinfo {author} {\bibfnamefont {F.~F.}\ \bibnamefont {Assaad}}, \
  and\ \bibinfo {author} {\bibfnamefont {M.}~\bibnamefont {Hohenadler}},\
  }\href {\doibase 10.1103/PhysRevB.94.155150} {\bibfield  {journal} {\bibinfo
  {journal} {Phys. Rev. B}\ }\textbf {\bibinfo {volume} {94}},\ \bibinfo
  {pages} {155150} (\bibinfo {year} {2016}{\natexlab{a}})}\BibitemShut
  {NoStop}%
\bibitem [{\citenamefont {Wei\ss{}e}\ and\ \citenamefont
  {Fehske}(1998)}]{PhysRevB.58.13526}%
  \BibitemOpen
  \bibfield  {author} {\bibinfo {author} {\bibfnamefont {A.}~\bibnamefont
  {Wei\ss{}e}}\ and\ \bibinfo {author} {\bibfnamefont {H.}~\bibnamefont
  {Fehske}},\ }\href {\doibase 10.1103/PhysRevB.58.13526} {\bibfield  {journal}
  {\bibinfo  {journal} {Phys. Rev. B}\ }\textbf {\bibinfo {volume} {58}},\
  \bibinfo {pages} {13526} (\bibinfo {year} {1998})}\BibitemShut {NoStop}%
\bibitem [{\citenamefont {Wellein}\ \emph {et~al.}(1998)\citenamefont
  {Wellein}, \citenamefont {Fehske},\ and\ \citenamefont
  {Kampf}}]{PhysRevLett.81.3956}%
  \BibitemOpen
  \bibfield  {author} {\bibinfo {author} {\bibfnamefont {G.}~\bibnamefont
  {Wellein}}, \bibinfo {author} {\bibfnamefont {H.}~\bibnamefont {Fehske}}, \
  and\ \bibinfo {author} {\bibfnamefont {A.~P.}\ \bibnamefont {Kampf}},\ }\href
  {\doibase 10.1103/PhysRevLett.81.3956} {\bibfield  {journal} {\bibinfo
  {journal} {Phys. Rev. Lett.}\ }\textbf {\bibinfo {volume} {81}},\ \bibinfo
  {pages} {3956} (\bibinfo {year} {1998})}\BibitemShut {NoStop}%
\bibitem [{\citenamefont {Hohenadler}\ \emph {et~al.}(2006)\citenamefont
  {Hohenadler}, \citenamefont {Wellein}, \citenamefont {Bishop}, \citenamefont
  {Alvermann},\ and\ \citenamefont {Fehske}}]{PhysRevB.73.245120}%
  \BibitemOpen
  \bibfield  {author} {\bibinfo {author} {\bibfnamefont {M.}~\bibnamefont
  {Hohenadler}}, \bibinfo {author} {\bibfnamefont {G.}~\bibnamefont {Wellein}},
  \bibinfo {author} {\bibfnamefont {A.~R.}\ \bibnamefont {Bishop}}, \bibinfo
  {author} {\bibfnamefont {A.}~\bibnamefont {Alvermann}}, \ and\ \bibinfo
  {author} {\bibfnamefont {H.}~\bibnamefont {Fehske}},\ }\href {\doibase
  10.1103/PhysRevB.73.245120} {\bibfield  {journal} {\bibinfo  {journal} {Phys.
  Rev. B}\ }\textbf {\bibinfo {volume} {73}},\ \bibinfo {pages} {245120}
  (\bibinfo {year} {2006})}\BibitemShut {NoStop}%
\bibitem [{\citenamefont {Fradkin}\ and\ \citenamefont
  {Hirsch}(1983)}]{PhysRevB.27.1680}%
  \BibitemOpen
  \bibfield  {author} {\bibinfo {author} {\bibfnamefont {E.}~\bibnamefont
  {Fradkin}}\ and\ \bibinfo {author} {\bibfnamefont {J.~E.}\ \bibnamefont
  {Hirsch}},\ }\href {\doibase 10.1103/PhysRevB.27.1680} {\bibfield  {journal}
  {\bibinfo  {journal} {Phys. Rev. B}\ }\textbf {\bibinfo {volume} {27}},\
  \bibinfo {pages} {1680} (\bibinfo {year} {1983})}\BibitemShut {NoStop}%
\bibitem [{\citenamefont {Hirsch}\ and\ \citenamefont
  {Fradkin}(1983)}]{PhysRevB.27.4302}%
  \BibitemOpen
  \bibfield  {author} {\bibinfo {author} {\bibfnamefont {J.~E.}\ \bibnamefont
  {Hirsch}}\ and\ \bibinfo {author} {\bibfnamefont {E.}~\bibnamefont
  {Fradkin}},\ }\href {\doibase 10.1103/PhysRevB.27.4302} {\bibfield  {journal}
  {\bibinfo  {journal} {Phys. Rev. B}\ }\textbf {\bibinfo {volume} {27}},\
  \bibinfo {pages} {4302} (\bibinfo {year} {1983})}\BibitemShut {NoStop}%
\bibitem [{\citenamefont {McKenzie}\ \emph {et~al.}(1996)\citenamefont
  {McKenzie}, \citenamefont {Hamer},\ and\ \citenamefont
  {Murray}}]{PhysRevB.53.9676}%
  \BibitemOpen
  \bibfield  {author} {\bibinfo {author} {\bibfnamefont {R.~H.}\ \bibnamefont
  {McKenzie}}, \bibinfo {author} {\bibfnamefont {C.~J.}\ \bibnamefont {Hamer}},
  \ and\ \bibinfo {author} {\bibfnamefont {D.~W.}\ \bibnamefont {Murray}},\
  }\href {\doibase 10.1103/PhysRevB.53.9676} {\bibfield  {journal} {\bibinfo
  {journal} {Phys. Rev. B}\ }\textbf {\bibinfo {volume} {53}},\ \bibinfo
  {pages} {9676} (\bibinfo {year} {1996})}\BibitemShut {NoStop}%
\bibitem [{\citenamefont {K\"uhne}\ and\ \citenamefont
  {L\"ow}(1999)}]{PhysRevB.60.12125}%
  \BibitemOpen
  \bibfield  {author} {\bibinfo {author} {\bibfnamefont {R.~W.}\ \bibnamefont
  {K\"uhne}}\ and\ \bibinfo {author} {\bibfnamefont {U.}~\bibnamefont
  {L\"ow}},\ }\href {\doibase 10.1103/PhysRevB.60.12125} {\bibfield  {journal}
  {\bibinfo  {journal} {Phys. Rev. B}\ }\textbf {\bibinfo {volume} {60}},\
  \bibinfo {pages} {12125} (\bibinfo {year} {1999})}\BibitemShut {NoStop}%
\bibitem [{\citenamefont {Sandvik}\ and\ \citenamefont
  {Campbell}(1999)}]{PhysRevLett.83.195}%
  \BibitemOpen
  \bibfield  {author} {\bibinfo {author} {\bibfnamefont {A.~W.}\ \bibnamefont
  {Sandvik}}\ and\ \bibinfo {author} {\bibfnamefont {D.~K.}\ \bibnamefont
  {Campbell}},\ }\href {\doibase 10.1103/PhysRevLett.83.195} {\bibfield
  {journal} {\bibinfo  {journal} {Phys. Rev. Lett.}\ }\textbf {\bibinfo
  {volume} {83}},\ \bibinfo {pages} {195} (\bibinfo {year} {1999})}\BibitemShut
  {NoStop}%
\bibitem [{\citenamefont {Hohenadler}\ \emph {et~al.}(2011)\citenamefont
  {Hohenadler}, \citenamefont {Fehske},\ and\ \citenamefont
  {Assaad}}]{PhysRevB.83.115105}%
  \BibitemOpen
  \bibfield  {author} {\bibinfo {author} {\bibfnamefont {M.}~\bibnamefont
  {Hohenadler}}, \bibinfo {author} {\bibfnamefont {H.}~\bibnamefont {Fehske}},
  \ and\ \bibinfo {author} {\bibfnamefont {F.~F.}\ \bibnamefont {Assaad}},\
  }\href {\doibase 10.1103/PhysRevB.83.115105} {\bibfield  {journal} {\bibinfo
  {journal} {Phys. Rev. B}\ }\textbf {\bibinfo {volume} {83}},\ \bibinfo
  {pages} {115105} (\bibinfo {year} {2011})}\BibitemShut {NoStop}%
\bibitem [{\citenamefont {Jeckelmann}\ \emph {et~al.}(1999)\citenamefont
  {Jeckelmann}, \citenamefont {Zhang},\ and\ \citenamefont
  {White}}]{PhysRevB.60.7950}%
  \BibitemOpen
  \bibfield  {author} {\bibinfo {author} {\bibfnamefont {E.}~\bibnamefont
  {Jeckelmann}}, \bibinfo {author} {\bibfnamefont {C.}~\bibnamefont {Zhang}}, \
  and\ \bibinfo {author} {\bibfnamefont {S.~R.}\ \bibnamefont {White}},\ }\href
  {\doibase 10.1103/PhysRevB.60.7950} {\bibfield  {journal} {\bibinfo
  {journal} {Phys. Rev. B}\ }\textbf {\bibinfo {volume} {60}},\ \bibinfo
  {pages} {7950} (\bibinfo {year} {1999})}\BibitemShut {NoStop}%
\bibitem [{\citenamefont {Barford}\ and\ \citenamefont
  {Bursill}(2005)}]{PhysRevLett.95.137207}%
  \BibitemOpen
  \bibfield  {author} {\bibinfo {author} {\bibfnamefont {W.}~\bibnamefont
  {Barford}}\ and\ \bibinfo {author} {\bibfnamefont {R.~J.}\ \bibnamefont
  {Bursill}},\ }\href {\doibase 10.1103/PhysRevLett.95.137207} {\bibfield
  {journal} {\bibinfo  {journal} {Phys. Rev. Lett.}\ }\textbf {\bibinfo
  {volume} {95}},\ \bibinfo {pages} {137207} (\bibinfo {year}
  {2005})}\BibitemShut {NoStop}%
\bibitem [{\citenamefont {Hager}\ \emph {et~al.}(2007)\citenamefont {Hager},
  \citenamefont {Wei{\ss}e}, \citenamefont {Wellein}, \citenamefont
  {Jeckelmann},\ and\ \citenamefont {Fehske}}]{Hager20071380}%
  \BibitemOpen
  \bibfield  {author} {\bibinfo {author} {\bibfnamefont {G.}~\bibnamefont
  {Hager}}, \bibinfo {author} {\bibfnamefont {A.}~\bibnamefont {Wei{\ss}e}},
  \bibinfo {author} {\bibfnamefont {G.}~\bibnamefont {Wellein}}, \bibinfo
  {author} {\bibfnamefont {E.}~\bibnamefont {Jeckelmann}}, \ and\ \bibinfo
  {author} {\bibfnamefont {H.}~\bibnamefont {Fehske}},\ }\href@noop {}
  {\bibfield  {journal} {\bibinfo  {journal} {J. Magn. Magn. Mater.}\ }\textbf
  {\bibinfo {volume} {310}},\ \bibinfo {pages} {1380 } (\bibinfo {year}
  {2007})}\BibitemShut {NoStop}%
\bibitem [{\citenamefont {Ejima}\ and\ \citenamefont
  {Fehske}(2009)}]{0295-5075-87-2-27001}%
  \BibitemOpen
  \bibfield  {author} {\bibinfo {author} {\bibfnamefont {S.}~\bibnamefont
  {Ejima}}\ and\ \bibinfo {author} {\bibfnamefont {H.}~\bibnamefont {Fehske}},\
  }\href {http://stacks.iop.org/0295-5075/87/i=2/a=27001} {\bibfield  {journal}
  {\bibinfo  {journal} {Europhys. Lett.}\ }\textbf {\bibinfo {volume} {87}},\
  \bibinfo {pages} {27001} (\bibinfo {year} {2009})}\BibitemShut {NoStop}%
\bibitem [{\citenamefont {Bursill}\ \emph {et~al.}(1998)\citenamefont
  {Bursill}, \citenamefont {McKenzie},\ and\ \citenamefont
  {Hamer}}]{PhysRevLett.80.5607}%
  \BibitemOpen
  \bibfield  {author} {\bibinfo {author} {\bibfnamefont {R.~J.}\ \bibnamefont
  {Bursill}}, \bibinfo {author} {\bibfnamefont {R.~H.}\ \bibnamefont
  {McKenzie}}, \ and\ \bibinfo {author} {\bibfnamefont {C.~J.}\ \bibnamefont
  {Hamer}},\ }\href {\doibase 10.1103/PhysRevLett.80.5607} {\bibfield
  {journal} {\bibinfo  {journal} {Phys. Rev. Lett.}\ }\textbf {\bibinfo
  {volume} {80}},\ \bibinfo {pages} {5607} (\bibinfo {year}
  {1998})}\BibitemShut {NoStop}%
\bibitem [{\citenamefont {Caron}\ and\ \citenamefont
  {Bourbonnais}(1984)}]{PhysRevB.29.4230}%
  \BibitemOpen
  \bibfield  {author} {\bibinfo {author} {\bibfnamefont {L.~G.}\ \bibnamefont
  {Caron}}\ and\ \bibinfo {author} {\bibfnamefont {C.}~\bibnamefont
  {Bourbonnais}},\ }\href {\doibase 10.1103/PhysRevB.29.4230} {\bibfield
  {journal} {\bibinfo  {journal} {Phys. Rev. B}\ }\textbf {\bibinfo {volume}
  {29}},\ \bibinfo {pages} {4230} (\bibinfo {year} {1984})}\BibitemShut
  {NoStop}%
\bibitem [{\citenamefont {Trebst}\ \emph {et~al.}(2001)\citenamefont {Trebst},
  \citenamefont {Elstner},\ and\ \citenamefont {Monien}}]{Trebst2001}%
  \BibitemOpen
  \bibfield  {author} {\bibinfo {author} {\bibfnamefont {S.}~\bibnamefont
  {Trebst}}, \bibinfo {author} {\bibfnamefont {N.}~\bibnamefont {Elstner}}, \
  and\ \bibinfo {author} {\bibfnamefont {H.}~\bibnamefont {Monien}},\ }\href
  {\doibase 10.1209/epl/i2001-00516-1} {\bibfield  {journal} {\bibinfo
  {journal} {Europhys. Lett.}\ }\textbf {\bibinfo {volume} {56}},\ \bibinfo
  {pages} {268} (\bibinfo {year} {2001})}\BibitemShut {NoStop}%
\bibitem [{\citenamefont {Sykora}\ \emph {et~al.}(2005)\citenamefont {Sykora},
  \citenamefont {H\"ubsch}, \citenamefont {Becker}, \citenamefont {Wellein},\
  and\ \citenamefont {Fehske}}]{PhysRevB.71.045112}%
  \BibitemOpen
  \bibfield  {author} {\bibinfo {author} {\bibfnamefont {S.}~\bibnamefont
  {Sykora}}, \bibinfo {author} {\bibfnamefont {A.}~\bibnamefont {H\"ubsch}},
  \bibinfo {author} {\bibfnamefont {K.~W.}\ \bibnamefont {Becker}}, \bibinfo
  {author} {\bibfnamefont {G.}~\bibnamefont {Wellein}}, \ and\ \bibinfo
  {author} {\bibfnamefont {H.}~\bibnamefont {Fehske}},\ }\href {\doibase
  10.1103/PhysRevB.71.045112} {\bibfield  {journal} {\bibinfo  {journal} {Phys.
  Rev. B}\ }\textbf {\bibinfo {volume} {71}},\ \bibinfo {pages} {045112}
  (\bibinfo {year} {2005})}\BibitemShut {NoStop}%
\bibitem [{\citenamefont {B\"uhler}\ \emph {et~al.}(2004)\citenamefont
  {B\"uhler}, \citenamefont {Uhrig},\ and\ \citenamefont
  {Oitmaa}}]{PhysRevB.70.214429}%
  \BibitemOpen
  \bibfield  {author} {\bibinfo {author} {\bibfnamefont {A.}~\bibnamefont
  {B\"uhler}}, \bibinfo {author} {\bibfnamefont {G.~S.}\ \bibnamefont {Uhrig}},
  \ and\ \bibinfo {author} {\bibfnamefont {J.}~\bibnamefont {Oitmaa}},\ }\href
  {\doibase 10.1103/PhysRevB.70.214429} {\bibfield  {journal} {\bibinfo
  {journal} {Phys. Rev. B}\ }\textbf {\bibinfo {volume} {70}},\ \bibinfo
  {pages} {214429} (\bibinfo {year} {2004})}\BibitemShut {NoStop}%
\bibitem [{\citenamefont {Bakrim}\ and\ \citenamefont
  {Bourbonnais}(2015)}]{Barkim2015}%
  \BibitemOpen
  \bibfield  {author} {\bibinfo {author} {\bibfnamefont {H.}~\bibnamefont
  {Bakrim}}\ and\ \bibinfo {author} {\bibfnamefont {C.}~\bibnamefont
  {Bourbonnais}},\ }\href {\doibase 10.1103/PhysRevB.91.085114} {\bibfield
  {journal} {\bibinfo  {journal} {Phys. Rev. B}\ }\textbf {\bibinfo {volume}
  {91}},\ \bibinfo {pages} {085114} (\bibinfo {year} {2015})}\BibitemShut
  {NoStop}%
\bibitem [{\citenamefont {Braden}\ \emph {et~al.}(1996)\citenamefont {Braden},
  \citenamefont {Wilkendorf}, \citenamefont {Lorenzana}, \citenamefont
  {A\"{\i}n}, \citenamefont {McIntyre}, \citenamefont {Behruzi}, \citenamefont
  {Heger}, \citenamefont {Dhalenne},\ and\ \citenamefont
  {Revcolevschi}}]{PhysRevB.54.1105}%
  \BibitemOpen
  \bibfield  {author} {\bibinfo {author} {\bibfnamefont {M.}~\bibnamefont
  {Braden}}, \bibinfo {author} {\bibfnamefont {G.}~\bibnamefont {Wilkendorf}},
  \bibinfo {author} {\bibfnamefont {J.}~\bibnamefont {Lorenzana}}, \bibinfo
  {author} {\bibfnamefont {M.}~\bibnamefont {A\"{\i}n}}, \bibinfo {author}
  {\bibfnamefont {G.~J.}\ \bibnamefont {McIntyre}}, \bibinfo {author}
  {\bibfnamefont {M.}~\bibnamefont {Behruzi}}, \bibinfo {author} {\bibfnamefont
  {G.}~\bibnamefont {Heger}}, \bibinfo {author} {\bibfnamefont
  {G.}~\bibnamefont {Dhalenne}}, \ and\ \bibinfo {author} {\bibfnamefont
  {A.}~\bibnamefont {Revcolevschi}},\ }\href {\doibase
  10.1103/PhysRevB.54.1105} {\bibfield  {journal} {\bibinfo  {journal} {Phys.
  Rev. B}\ }\textbf {\bibinfo {volume} {54}},\ \bibinfo {pages} {1105}
  (\bibinfo {year} {1996})}\BibitemShut {NoStop}%
\bibitem [{\citenamefont {Feiguin}\ and\ \citenamefont
  {Fiete}(2010)}]{PhysRevB.81.075108}%
  \BibitemOpen
  \bibfield  {author} {\bibinfo {author} {\bibfnamefont {A.~E.}\ \bibnamefont
  {Feiguin}}\ and\ \bibinfo {author} {\bibfnamefont {G.~A.}\ \bibnamefont
  {Fiete}},\ }\href {\doibase 10.1103/PhysRevB.81.075108} {\bibfield  {journal}
  {\bibinfo  {journal} {Phys. Rev. B}\ }\textbf {\bibinfo {volume} {81}},\
  \bibinfo {pages} {075108} (\bibinfo {year} {2010})}\BibitemShut {NoStop}%
\bibitem [{\citenamefont {Karrasch}\ and\ \citenamefont
  {Moore}(2012)}]{PhysRevB.86.155156}%
  \BibitemOpen
  \bibfield  {author} {\bibinfo {author} {\bibfnamefont {C.}~\bibnamefont
  {Karrasch}}\ and\ \bibinfo {author} {\bibfnamefont {J.~E.}\ \bibnamefont
  {Moore}},\ }\href {\doibase 10.1103/PhysRevB.86.155156} {\bibfield  {journal}
  {\bibinfo  {journal} {Phys. Rev. B}\ }\textbf {\bibinfo {volume} {86}},\
  \bibinfo {pages} {155156} (\bibinfo {year} {2012})}\BibitemShut {NoStop}%
\bibitem [{\citenamefont {Hohenadler}\ and\ \citenamefont
  {Lang}(2008)}]{Hohenadler2008}%
  \BibitemOpen
  \bibfield  {author} {\bibinfo {author} {\bibfnamefont {M.}~\bibnamefont
  {Hohenadler}}\ and\ \bibinfo {author} {\bibfnamefont {T.~C.}\ \bibnamefont
  {Lang}},\ }in\ \href {\doibase 10.1007/978-3-540-74686-7_11} {\emph {\bibinfo
  {booktitle} {Computational Many-Particle Physics}}},\ \bibinfo {editor}
  {edited by\ \bibinfo {editor} {\bibfnamefont {H.}~\bibnamefont {Fehske}},
  \bibinfo {editor} {\bibfnamefont {R.}~\bibnamefont {Schneider}}, \ and\
  \bibinfo {editor} {\bibfnamefont {A.}~\bibnamefont {Wei{\ss}e}}}\ (\bibinfo
  {publisher} {Springer Berlin Heidelberg},\ \bibinfo {address} {Berlin,
  Heidelberg},\ \bibinfo {year} {2008})\ pp.\ \bibinfo {pages}
  {357--366}\BibitemShut {NoStop}%
\bibitem [{\citenamefont {Fye}\ and\ \citenamefont
  {Scalettar}(1987)}]{PhysRevB.36.3833}%
  \BibitemOpen
  \bibfield  {author} {\bibinfo {author} {\bibfnamefont {R.~M.}\ \bibnamefont
  {Fye}}\ and\ \bibinfo {author} {\bibfnamefont {R.~T.}\ \bibnamefont
  {Scalettar}},\ }\href {\doibase 10.1103/PhysRevB.36.3833} {\bibfield
  {journal} {\bibinfo  {journal} {Phys. Rev. B}\ }\textbf {\bibinfo {volume}
  {36}},\ \bibinfo {pages} {3833} (\bibinfo {year} {1987})}\BibitemShut
  {NoStop}%
\bibitem [{\citenamefont {Orignac}\ and\ \citenamefont
  {Chitra}(2004)}]{PhysRevB.70.214436}%
  \BibitemOpen
  \bibfield  {author} {\bibinfo {author} {\bibfnamefont {E.}~\bibnamefont
  {Orignac}}\ and\ \bibinfo {author} {\bibfnamefont {R.}~\bibnamefont
  {Chitra}},\ }\href {\doibase 10.1103/PhysRevB.70.214436} {\bibfield
  {journal} {\bibinfo  {journal} {Phys. Rev. B}\ }\textbf {\bibinfo {volume}
  {70}},\ \bibinfo {pages} {214436} (\bibinfo {year} {2004})}\BibitemShut
  {NoStop}%
\bibitem [{\citenamefont {Voit}\ and\ \citenamefont
  {Schulz}(1987)}]{PhysRevB.36.968}%
  \BibitemOpen
  \bibfield  {author} {\bibinfo {author} {\bibfnamefont {J.}~\bibnamefont
  {Voit}}\ and\ \bibinfo {author} {\bibfnamefont {H.~J.}\ \bibnamefont
  {Schulz}},\ }\href {\doibase 10.1103/PhysRevB.36.968} {\bibfield  {journal}
  {\bibinfo  {journal} {Phys. Rev. B}\ }\textbf {\bibinfo {volume} {36}},\
  \bibinfo {pages} {968} (\bibinfo {year} {1987})}\BibitemShut {NoStop}%
\bibitem [{\citenamefont {Weber}\ \emph {et~al.}(2017)\citenamefont {Weber},
  \citenamefont {Assaad},\ and\ \citenamefont
  {Hohenadler}}]{PhysRevLett.119.097401}%
  \BibitemOpen
  \bibfield  {author} {\bibinfo {author} {\bibfnamefont {M.}~\bibnamefont
  {Weber}}, \bibinfo {author} {\bibfnamefont {F.~F.}\ \bibnamefont {Assaad}}, \
  and\ \bibinfo {author} {\bibfnamefont {M.}~\bibnamefont {Hohenadler}},\
  }\href {\doibase 10.1103/PhysRevLett.119.097401} {\bibfield  {journal}
  {\bibinfo  {journal} {Phys. Rev. Lett.}\ }\textbf {\bibinfo {volume} {119}},\
  \bibinfo {pages} {097401} (\bibinfo {year} {2017})}\BibitemShut {NoStop}%
\bibitem [{\citenamefont {Weber}\ \emph
  {et~al.}(2016{\natexlab{b}})\citenamefont {Weber}, \citenamefont {Assaad},\
  and\ \citenamefont {Hohenadler}}]{PhysRevB.94.245138}%
  \BibitemOpen
  \bibfield  {author} {\bibinfo {author} {\bibfnamefont {M.}~\bibnamefont
  {Weber}}, \bibinfo {author} {\bibfnamefont {F.~F.}\ \bibnamefont {Assaad}}, \
  and\ \bibinfo {author} {\bibfnamefont {M.}~\bibnamefont {Hohenadler}},\
  }\href {\doibase 10.1103/PhysRevB.94.245138} {\bibfield  {journal} {\bibinfo
  {journal} {Phys. Rev. B}\ }\textbf {\bibinfo {volume} {94}},\ \bibinfo
  {pages} {245138} (\bibinfo {year} {2016}{\natexlab{b}})}\BibitemShut
  {NoStop}%
\bibitem [{\citenamefont {Holstein}(1959)}]{Ho59a}%
  \BibitemOpen
  \bibfield  {author} {\bibinfo {author} {\bibfnamefont {T.}~\bibnamefont
  {Holstein}},\ }\href@noop {} {\bibfield  {journal} {\bibinfo  {journal} {Ann.
  Phys. (N.Y.)}\ }\textbf {\bibinfo {volume} {8}},\ \bibinfo {pages} {325
  (1959); {\bf 8}, 343} (\bibinfo {year} {1959})}\BibitemShut {NoStop}%
\bibitem [{\citenamefont {Hohenadler}\ and\ \citenamefont
  {Fehske}(2017)}]{MHHF2017}%
  \BibitemOpen
  \bibfield  {author} {\bibinfo {author} {\bibfnamefont {M.}~\bibnamefont
  {Hohenadler}}\ and\ \bibinfo {author} {\bibfnamefont {H.}~\bibnamefont
  {Fehske}},\ }\href@noop {} {\bibfield  {journal} {\bibinfo  {journal}
  {arXiv:1706.00470}\ } (\bibinfo {year} {2017})}\BibitemShut {NoStop}%
\bibitem [{\citenamefont {Meden}\ \emph {et~al.}(1994)\citenamefont {Meden},
  \citenamefont {Sch\"onhammer},\ and\ \citenamefont
  {Gunnarsson}}]{PhysRevB.50.11179}%
  \BibitemOpen
  \bibfield  {author} {\bibinfo {author} {\bibfnamefont {V.}~\bibnamefont
  {Meden}}, \bibinfo {author} {\bibfnamefont {K.}~\bibnamefont
  {Sch\"onhammer}}, \ and\ \bibinfo {author} {\bibfnamefont {O.}~\bibnamefont
  {Gunnarsson}},\ }\href {\doibase 10.1103/PhysRevB.50.11179} {\bibfield
  {journal} {\bibinfo  {journal} {Phys. Rev. B}\ }\textbf {\bibinfo {volume}
  {50}},\ \bibinfo {pages} {11179} (\bibinfo {year} {1994})}\BibitemShut
  {NoStop}%
\bibitem [{\citenamefont {Sykora}\ \emph {et~al.}(2006)\citenamefont {Sykora},
  \citenamefont {H\"ubsch},\ and\ \citenamefont {Becker}}]{0295-5075-76-4-644}%
  \BibitemOpen
  \bibfield  {author} {\bibinfo {author} {\bibfnamefont {S.}~\bibnamefont
  {Sykora}}, \bibinfo {author} {\bibfnamefont {A.}~\bibnamefont {H\"ubsch}}, \
  and\ \bibinfo {author} {\bibfnamefont {K.~W.}\ \bibnamefont {Becker}},\
  }\href {http://stacks.iop.org/0295-5075/76/i=4/a=644} {\bibfield  {journal}
  {\bibinfo  {journal} {Europhys. Lett.}\ }\textbf {\bibinfo {volume} {76}},\
  \bibinfo {pages} {644} (\bibinfo {year} {2006})}\BibitemShut {NoStop}%
\bibitem [{\citenamefont {Greitemann}\ \emph {et~al.}(2015)\citenamefont
  {Greitemann}, \citenamefont {Hesselmann}, \citenamefont {Wessel},
  \citenamefont {Assaad},\ and\ \citenamefont
  {Hohenadler}}]{PhysRevB.92.245132}%
  \BibitemOpen
  \bibfield  {author} {\bibinfo {author} {\bibfnamefont {J.}~\bibnamefont
  {Greitemann}}, \bibinfo {author} {\bibfnamefont {S.}~\bibnamefont
  {Hesselmann}}, \bibinfo {author} {\bibfnamefont {S.}~\bibnamefont {Wessel}},
  \bibinfo {author} {\bibfnamefont {F.~F.}\ \bibnamefont {Assaad}}, \ and\
  \bibinfo {author} {\bibfnamefont {M.}~\bibnamefont {Hohenadler}},\ }\href
  {\doibase 10.1103/PhysRevB.92.245132} {\bibfield  {journal} {\bibinfo
  {journal} {Phys. Rev. B}\ }\textbf {\bibinfo {volume} {92}},\ \bibinfo
  {pages} {245132} (\bibinfo {year} {2015})}\BibitemShut {NoStop}%
\bibitem [{\citenamefont {Weber}\ \emph
  {et~al.}(2015{\natexlab{a}})\citenamefont {Weber}, \citenamefont {Assaad},\
  and\ \citenamefont {Hohenadler}}]{PhysRevB.91.245147}%
  \BibitemOpen
  \bibfield  {author} {\bibinfo {author} {\bibfnamefont {M.}~\bibnamefont
  {Weber}}, \bibinfo {author} {\bibfnamefont {F.~F.}\ \bibnamefont {Assaad}}, \
  and\ \bibinfo {author} {\bibfnamefont {M.}~\bibnamefont {Hohenadler}},\
  }\href {\doibase 10.1103/PhysRevB.91.245147} {\bibfield  {journal} {\bibinfo
  {journal} {Phys. Rev. B}\ }\textbf {\bibinfo {volume} {91}},\ \bibinfo
  {pages} {245147} (\bibinfo {year} {2015}{\natexlab{a}})}\BibitemShut
  {NoStop}%
\bibitem [{\citenamefont {Cross}\ and\ \citenamefont
  {Fisher}(1979)}]{PhysRevB.19.402}%
  \BibitemOpen
  \bibfield  {author} {\bibinfo {author} {\bibfnamefont {M.~C.}\ \bibnamefont
  {Cross}}\ and\ \bibinfo {author} {\bibfnamefont {D.~S.}\ \bibnamefont
  {Fisher}},\ }\href {\doibase 10.1103/PhysRevB.19.402} {\bibfield  {journal}
  {\bibinfo  {journal} {Phys. Rev. B}\ }\textbf {\bibinfo {volume} {19}},\
  \bibinfo {pages} {402} (\bibinfo {year} {1979})}\BibitemShut {NoStop}%
\bibitem [{\citenamefont {Werner}\ \emph {et~al.}(1999)\citenamefont {Werner},
  \citenamefont {Gros},\ and\ \citenamefont {Braden}}]{PhysRevB.59.14356}%
  \BibitemOpen
  \bibfield  {author} {\bibinfo {author} {\bibfnamefont {R.}~\bibnamefont
  {Werner}}, \bibinfo {author} {\bibfnamefont {C.}~\bibnamefont {Gros}}, \ and\
  \bibinfo {author} {\bibfnamefont {M.}~\bibnamefont {Braden}},\ }\href
  {\doibase 10.1103/PhysRevB.59.14356} {\bibfield  {journal} {\bibinfo
  {journal} {Phys. Rev. B}\ }\textbf {\bibinfo {volume} {59}},\ \bibinfo
  {pages} {14356} (\bibinfo {year} {1999})}\BibitemShut {NoStop}%
\bibitem [{\citenamefont {Zimanyi}\ \emph {et~al.}(1988)\citenamefont
  {Zimanyi}, \citenamefont {Kivelson},\ and\ \citenamefont
  {Luther}}]{PhysRevLett.60.2089}%
  \BibitemOpen
  \bibfield  {author} {\bibinfo {author} {\bibfnamefont {G.~T.}\ \bibnamefont
  {Zimanyi}}, \bibinfo {author} {\bibfnamefont {S.~A.}\ \bibnamefont
  {Kivelson}}, \ and\ \bibinfo {author} {\bibfnamefont {A.}~\bibnamefont
  {Luther}},\ }\href {\doibase 10.1103/PhysRevLett.60.2089} {\bibfield
  {journal} {\bibinfo  {journal} {Phys. Rev. Lett.}\ }\textbf {\bibinfo
  {volume} {60}},\ \bibinfo {pages} {2089} (\bibinfo {year}
  {1988})}\BibitemShut {NoStop}%
\bibitem [{\citenamefont {Feynman}(1955)}]{PhysRev.97.660}%
  \BibitemOpen
  \bibfield  {author} {\bibinfo {author} {\bibfnamefont {R.~P.}\ \bibnamefont
  {Feynman}},\ }\href {\doibase 10.1103/PhysRev.97.660} {\bibfield  {journal}
  {\bibinfo  {journal} {Phys. Rev.}\ }\textbf {\bibinfo {volume} {97}},\
  \bibinfo {pages} {660} (\bibinfo {year} {1955})}\BibitemShut {NoStop}%
\bibitem [{\citenamefont {Sandvik}\ and\ \citenamefont
  {Kurkij\"arvi}(1991)}]{PhysRevB.43.5950}%
  \BibitemOpen
  \bibfield  {author} {\bibinfo {author} {\bibfnamefont {A.~W.}\ \bibnamefont
  {Sandvik}}\ and\ \bibinfo {author} {\bibfnamefont {J.}~\bibnamefont
  {Kurkij\"arvi}},\ }\href {\doibase 10.1103/PhysRevB.43.5950} {\bibfield
  {journal} {\bibinfo  {journal} {Phys. Rev. B}\ }\textbf {\bibinfo {volume}
  {43}},\ \bibinfo {pages} {5950} (\bibinfo {year} {1991})}\BibitemShut
  {NoStop}%
\bibitem [{\citenamefont {Syljuasen}\ and\ \citenamefont
  {Sandvik}(2002)}]{Sandvik02}%
  \BibitemOpen
  \bibfield  {author} {\bibinfo {author} {\bibfnamefont {O.}~\bibnamefont
  {Syljuasen}}\ and\ \bibinfo {author} {\bibfnamefont {A.~W.}\ \bibnamefont
  {Sandvik}},\ }\href@noop {} {\bibfield  {journal} {\bibinfo  {journal} {Phys.
  Rev. E}\ }\textbf {\bibinfo {volume} {66}},\ \bibinfo {pages} {046701}
  (\bibinfo {year} {2002})}\BibitemShut {NoStop}%
\bibitem [{\citenamefont {Sandvik}(1992)}]{Sandvik91}%
  \BibitemOpen
  \bibfield  {author} {\bibinfo {author} {\bibfnamefont {A.~W.}\ \bibnamefont
  {Sandvik}},\ }\href {http://stacks.iop.org/0305-4470/25/i=13/a=017}
  {\bibfield  {journal} {\bibinfo  {journal} {J. Phys. A: Math. Gen.}\ }\textbf
  {\bibinfo {volume} {25}},\ \bibinfo {pages} {3667} (\bibinfo {year}
  {1992})}\BibitemShut {NoStop}%
\bibitem [{\citenamefont {Sandvik}\ \emph {et~al.}(1997)\citenamefont
  {Sandvik}, \citenamefont {Singh},\ and\ \citenamefont
  {Campbell}}]{PhysRevB.56.14510}%
  \BibitemOpen
  \bibfield  {author} {\bibinfo {author} {\bibfnamefont {A.~W.}\ \bibnamefont
  {Sandvik}}, \bibinfo {author} {\bibfnamefont {R.~R.~P.}\ \bibnamefont
  {Singh}}, \ and\ \bibinfo {author} {\bibfnamefont {D.~K.}\ \bibnamefont
  {Campbell}},\ }\href {\doibase 10.1103/PhysRevB.56.14510} {\bibfield
  {journal} {\bibinfo  {journal} {Phys. Rev. B}\ }\textbf {\bibinfo {volume}
  {56}},\ \bibinfo {pages} {14510} (\bibinfo {year} {1997})}\BibitemShut
  {NoStop}%
\bibitem [{\citenamefont {Huscroft}\ \emph {et~al.}(2000)\citenamefont
  {Huscroft}, \citenamefont {Gass},\ and\ \citenamefont
  {Jarrell}}]{PhysRevB.61.9300}%
  \BibitemOpen
  \bibfield  {author} {\bibinfo {author} {\bibfnamefont {C.}~\bibnamefont
  {Huscroft}}, \bibinfo {author} {\bibfnamefont {R.}~\bibnamefont {Gass}}, \
  and\ \bibinfo {author} {\bibfnamefont {M.}~\bibnamefont {Jarrell}},\ }\href
  {\doibase 10.1103/PhysRevB.61.9300} {\bibfield  {journal} {\bibinfo
  {journal} {Phys. Rev. B}\ }\textbf {\bibinfo {volume} {61}},\ \bibinfo
  {pages} {9300} (\bibinfo {year} {2000})}\BibitemShut {NoStop}%
\bibitem [{\citenamefont {{Jarrell}}\ and\ \citenamefont
  {{Gubernatis}}(1996)}]{Jarrell1996133}%
  \BibitemOpen
  \bibfield  {author} {\bibinfo {author} {\bibfnamefont {M.}~\bibnamefont
  {{Jarrell}}}\ and\ \bibinfo {author} {\bibfnamefont {J.~E.}\ \bibnamefont
  {{Gubernatis}}},\ }\href {\doibase 10.1016/0370-1573(95)00074-7} {\bibfield
  {journal} {\bibinfo  {journal} {Phys. Rep.}\ }\textbf {\bibinfo {volume}
  {269}},\ \bibinfo {pages} {133} (\bibinfo {year} {1996})}\BibitemShut
  {NoStop}%
\bibitem [{\citenamefont {Dorneich}\ and\ \citenamefont
  {Troyer}(2001)}]{PhysRevE.64.066701}%
  \BibitemOpen
  \bibfield  {author} {\bibinfo {author} {\bibfnamefont {A.}~\bibnamefont
  {Dorneich}}\ and\ \bibinfo {author} {\bibfnamefont {M.}~\bibnamefont
  {Troyer}},\ }\href {\doibase 10.1103/PhysRevE.64.066701} {\bibfield
  {journal} {\bibinfo  {journal} {Phys. Rev. E}\ }\textbf {\bibinfo {volume}
  {64}},\ \bibinfo {pages} {066701} (\bibinfo {year} {2001})}\BibitemShut
  {NoStop}%
\bibitem [{\citenamefont {Weber}\ \emph
  {et~al.}(2015{\natexlab{b}})\citenamefont {Weber}, \citenamefont {Assaad},\
  and\ \citenamefont {Hohenadler}}]{PhysRevB.91.235150}%
  \BibitemOpen
  \bibfield  {author} {\bibinfo {author} {\bibfnamefont {M.}~\bibnamefont
  {Weber}}, \bibinfo {author} {\bibfnamefont {F.~F.}\ \bibnamefont {Assaad}}, \
  and\ \bibinfo {author} {\bibfnamefont {M.}~\bibnamefont {Hohenadler}},\
  }\href {\doibase 10.1103/PhysRevB.91.235150} {\bibfield  {journal} {\bibinfo
  {journal} {Phys. Rev. B}\ }\textbf {\bibinfo {volume} {91}},\ \bibinfo
  {pages} {235150} (\bibinfo {year} {2015}{\natexlab{b}})}\BibitemShut
  {NoStop}%
\bibitem [{\citenamefont {Sandvik}(1998)}]{PhysRevB.57.10287}%
  \BibitemOpen
  \bibfield  {author} {\bibinfo {author} {\bibfnamefont {A.~W.}\ \bibnamefont
  {Sandvik}},\ }\href {\doibase 10.1103/PhysRevB.57.10287} {\bibfield
  {journal} {\bibinfo  {journal} {Phys. Rev. B}\ }\textbf {\bibinfo {volume}
  {57}},\ \bibinfo {pages} {10287} (\bibinfo {year} {1998})}\BibitemShut
  {NoStop}%
\bibitem [{\citenamefont {{Beach}}(2004)}]{2004cond.mat..3055B}%
  \BibitemOpen
  \bibfield  {author} {\bibinfo {author} {\bibfnamefont {K.~S.~D.}\
  \bibnamefont {{Beach}}},\ }\href@noop {} {\bibfield  {journal} {\bibinfo
  {journal} {arXiv:cond-mat/0403055}\ } (\bibinfo {year} {2004})}\BibitemShut
  {NoStop}%
\bibitem [{\citenamefont {Kadanoff}\ and\ \citenamefont
  {Baym}(1989)}]{KadanoffBaym}%
  \BibitemOpen
  \bibfield  {author} {\bibinfo {author} {\bibfnamefont {L.~P.}\ \bibnamefont
  {Kadanoff}}\ and\ \bibinfo {author} {\bibfnamefont {G.~A.}\ \bibnamefont
  {Baym}},\ }\href@noop {} {\emph {\bibinfo {title} {Quantum Statistical
  Mechanics: Green's Function Methods in Equilibrium and Nonequilibrium
  Problems}}}\ (\bibinfo  {publisher} {Addison-Wesley},\ \bibinfo {address}
  {Redwood City, California},\ \bibinfo {year} {1989})\BibitemShut {NoStop}%
\bibitem [{\citenamefont {Schneider}\ and\ \citenamefont
  {Wierstorf}(2014)}]{anna_schneider_2014_10282}%
  \BibitemOpen
  \bibfield  {author} {\bibinfo {author} {\bibfnamefont {A.}~\bibnamefont
  {Schneider}}\ and\ \bibinfo {author} {\bibfnamefont {H.}~\bibnamefont
  {Wierstorf}},\ }\href {\doibase 10.5281/zenodo.10282} {\enquote {\bibinfo
  {title} {{Gnuplot-colorbrewer: ColorBrewer color schemes for gnuplot}},}\
  }\bibinfo {howpublished} {10.5281/zenodo.10282} (\bibinfo {year}
  {2014})\BibitemShut {NoStop}%
\bibitem [{\citenamefont {Lang}\ and\ \citenamefont
  {Firsov}(1962)}]{LangFirsov}%
  \BibitemOpen
  \bibfield  {author} {\bibinfo {author} {\bibfnamefont {I.~G.}\ \bibnamefont
  {Lang}}\ and\ \bibinfo {author} {\bibfnamefont {Y.~A.}\ \bibnamefont
  {Firsov}},\ }\href@noop {} {\bibfield  {journal} {\bibinfo  {journal} {Zh.
  Eksp. Teor. Fiz.}\ }\textbf {\bibinfo {volume} {43}},\ \bibinfo {pages}
  {1843} (\bibinfo {year} {1962})},\ \bibinfo {note} {[Sov. Phys. JETP {\bf
  16}, 1301 (1962)]}\BibitemShut {NoStop}%
\bibitem [{\citenamefont {Loos}\ \emph {et~al.}(2006)\citenamefont {Loos},
  \citenamefont {Hohenadler},\ and\ \citenamefont
  {Fehske}}]{0953-8984-18-8-011}%
  \BibitemOpen
  \bibfield  {author} {\bibinfo {author} {\bibfnamefont {J.}~\bibnamefont
  {Loos}}, \bibinfo {author} {\bibfnamefont {M.}~\bibnamefont {Hohenadler}}, \
  and\ \bibinfo {author} {\bibfnamefont {H.}~\bibnamefont {Fehske}},\ }\href
  {http://stacks.iop.org/0953-8984/18/i=8/a=011} {\bibfield  {journal}
  {\bibinfo  {journal} {J. Phys.: Condens. Matter}\ }\textbf {\bibinfo {volume}
  {18}},\ \bibinfo {pages} {2453} (\bibinfo {year} {2006})}\BibitemShut
  {NoStop}%
\bibitem [{\citenamefont {Giamarchi}(2004)}]{Giamarchi:743140}%
  \BibitemOpen
  \bibfield  {author} {\bibinfo {author} {\bibfnamefont {T.}~\bibnamefont
  {Giamarchi}},\ }\href {https://cds.cern.ch/record/743140} {\emph {\bibinfo
  {title} {{Quantum physics in one dimension}}}},\ Internat. Ser. Mono. Phys.\
  (\bibinfo  {publisher} {Clarendon Press},\ \bibinfo {address} {Oxford},\
  \bibinfo {year} {2004})\BibitemShut {NoStop}%
\bibitem [{\citenamefont {Voit}\ \emph {et~al.}(2000)\citenamefont {Voit},
  \citenamefont {Perfetti}, \citenamefont {Zwick}, \citenamefont {Berger},
  \citenamefont {Margaritondo}, \citenamefont {Gr{\"u}ner}, \citenamefont
  {H{\"o}chst},\ and\ \citenamefont {Grioni}}]{Voit501}%
  \BibitemOpen
  \bibfield  {author} {\bibinfo {author} {\bibfnamefont {J.}~\bibnamefont
  {Voit}}, \bibinfo {author} {\bibfnamefont {L.}~\bibnamefont {Perfetti}},
  \bibinfo {author} {\bibfnamefont {F.}~\bibnamefont {Zwick}}, \bibinfo
  {author} {\bibfnamefont {H.}~\bibnamefont {Berger}}, \bibinfo {author}
  {\bibfnamefont {G.}~\bibnamefont {Margaritondo}}, \bibinfo {author}
  {\bibfnamefont {G.}~\bibnamefont {Gr{\"u}ner}}, \bibinfo {author}
  {\bibfnamefont {H.}~\bibnamefont {H{\"o}chst}}, \ and\ \bibinfo {author}
  {\bibfnamefont {M.}~\bibnamefont {Grioni}},\ }\href {\doibase
  10.1126/science.290.5491.501} {\bibfield  {journal} {\bibinfo  {journal}
  {Science}\ }\textbf {\bibinfo {volume} {290}},\ \bibinfo {pages} {501}
  (\bibinfo {year} {2000})}\BibitemShut {NoStop}%
\bibitem [{\citenamefont {{Michielsen}}\ and\ \citenamefont {{de
  Raedt}}(1996)}]{1996MPLB...10..467M}%
  \BibitemOpen
  \bibfield  {author} {\bibinfo {author} {\bibfnamefont {K.}~\bibnamefont
  {{Michielsen}}}\ and\ \bibinfo {author} {\bibfnamefont {H.}~\bibnamefont {{de
  Raedt}}},\ }\href {\doibase 10.1142/S0217984996000511} {\bibfield  {journal}
  {\bibinfo  {journal} {Mod. Phys. Lett. B}\ }\textbf {\bibinfo {volume}
  {10}},\ \bibinfo {pages} {467} (\bibinfo {year} {1996})}\BibitemShut
  {NoStop}%
\bibitem [{\citenamefont {Michel}\ and\ \citenamefont
  {Evertz}(2007)}]{arXiv:0705.0799}%
  \BibitemOpen
  \bibfield  {author} {\bibinfo {author} {\bibfnamefont {F.}~\bibnamefont
  {Michel}}\ and\ \bibinfo {author} {\bibfnamefont {H.~G.}\ \bibnamefont
  {Evertz}},\ }\href@noop {} {\bibfield  {journal} {\bibinfo  {journal}
  {arXiv:0705.0799}\ } (\bibinfo {year} {2007})}\BibitemShut {NoStop}%
\bibitem [{\citenamefont {Assaad}\ and\ \citenamefont
  {Evertz}(2008)}]{Assaad2008}%
  \BibitemOpen
  \bibfield  {author} {\bibinfo {author} {\bibfnamefont {F.~F.}\ \bibnamefont
  {Assaad}}\ and\ \bibinfo {author} {\bibfnamefont {H.~G.}\ \bibnamefont
  {Evertz}},\ }in\ \href {\doibase 10.1007/978-3-540-74686-7_10} {\emph
  {\bibinfo {booktitle} {Computational Many-Particle Physics}}},\ \bibinfo
  {editor} {edited by\ \bibinfo {editor} {\bibfnamefont {H.}~\bibnamefont
  {Fehske}}, \bibinfo {editor} {\bibfnamefont {R.}~\bibnamefont {Schneider}}, \
  and\ \bibinfo {editor} {\bibfnamefont {A.}~\bibnamefont {Wei{\ss}e}}}\
  (\bibinfo  {publisher} {Springer Berlin Heidelberg},\ \bibinfo {address}
  {Berlin, Heidelberg},\ \bibinfo {year} {2008})\ pp.\ \bibinfo {pages}
  {277--356}\BibitemShut {NoStop}%
\bibitem [{\citenamefont {{J\"ulich Supercomputing Centre}}(2016)}]{Juelich}%
  \BibitemOpen
  \bibfield  {author} {\bibinfo {author} {\bibnamefont {{J\"ulich
  Supercomputing Centre}}},\ }\href@noop {} {\bibfield  {journal} {\bibinfo
  {journal} {J. Large-Scale Res. Facilities}\ }\textbf {\bibinfo {volume}
  {2}},\ \bibinfo {pages} {A62} (\bibinfo {year} {2016})}\BibitemShut {NoStop}%
\end{thebibliography}

%


\end{document}